# Characteristics of stratified flows of Newtonian/non-Newtonian shear-thinning fluids

Davide Picchi[a*], Pietro Poesio[b], Amos Ullmann[a] and Neima Brauner[a]

[a] *Tel-Aviv University, Faculty of Engineering, School of Mechanical Engineering*
*Ramat Aviv, Tel-Aviv, 69978, Israel*
[b] *Università degli Studi di Brescia, Dipartimento di Ingegneria Meccanica ed Industriale,*
*Via Branze 38, 25123,Brescia, Italy*
*davide.picchi@gmail.com , pietro.poesio@unibs.it, ullmann@eng.tau.ac.il, brauner@eng.tau.ac.il*

**Abstract**

Exact solutions for laminar stratified flows of Newtonian/non-Newtonian shear-thinning fluids in horizontal and inclined channels are presented. An iterative algorithm is proposed to compute the laminar solution for the general case of a Carreau non-Newtonian fluid. The exact solution is used to study the effect of the rheology of the shear-thinning liquid on two-phase flow characteristics considering both gas/liquid and liquid/liquid systems. Concurrent and counter-current inclined systems are investigated, including the mapping of multiple solution boundaries. Aspects relevant to practical applications are discussed, such as the insitu hold-up, or lubrication effects achieved by adding a less viscous phase. A characteristic of this family of systems is that, even if the liquid has a complex rheology (Carreau fluid), the two-phase stratified flow can behave like the liquid is Newtonian for a wide range of operational conditions. The capability of the two-fluid model to yield satisfactory predictions in the presence of shear-thinning liquids is tested, and an algorithm is proposed to a priori predict if the Newtonian (zero shear rate viscosity) behaviour arises for a given operational conditions in order to avoid large errors in the predictions of flow characteristics when the power-law is considered for modelling the shear-thinning behaviour. Two-fluid model closures implied by the exact solution and the effect of a turbulent gas layer are also addressed.

## 1. Introduction

The stratified flow regime is the basic flow pattern in horizontal or slightly inclined gas-liquid and liquid-liquid systems in a gravity field, and it can be encountered both in concurrent and counter-current configurations. The stratified flow, where one of the two phases is a non-Newtonian shear-thinning fluid, is frequently encountered in several industrial applications, such as petroleum transport in pipelines, cleaning processes in food and chemical industry and polymer extrusion.

Complex fluids, such as dense emulsions, crude oils, waxy oils (slurries flowing with wax deposits), foams, high molecular fluids (melt polymer, polymer solutions, and proteins), and pharmaceuticals, behave as shear thinning fluids. For this family of non-Newtonian fluids, the effective viscosity is a function of the imposed shear rate and it has to be described by an appropriate rheological model. The Carreau (1972) viscosity model is known to be a good approximation for a large number of shear-thinning fluids due to its capability to catch the rheological behaviour at very low and very high shear rates, while other idealized models, like the widely used Ostwald de Waele power-law model, are valid only in a limited range of operational conditions.

Despite its importance for industrial applications, the issue of how the rheology of a shear-thinning fluid affects the characteristics of two-phase stratified flow at different flow conditions is not yet completely understood. Only few works have been published on stratified flows in the presence of shear-thinning fluids. Yu and Dae Han (1973), for example, studied the horizontal stratified flow of molten polymers, which were modelled as power-law fluids. Some works analysed the flow instabilities of a thin non-Newtonian layer (e.g. Khomami (1990a, b), Pinarbasi and Liakopoulos (1995), Sahu *et al.* (2007), Ó Náraigh and Spelt (2010), and Alba *et al.* (2013)). Firouzi and Hashemabadi (2009) presented the laminar solution for a Newtonian-fluid flowing over a thin non-Newtonian layer, but the study was limited to the case of a horizontal flow and a positive shear rate in the non-Newtonian layer. To the best of our knowledge, an exact solution for the general case of laminar stratified concurrent and counter-current flows in horizontal and inclined channels is not yet available the literature.

Recently, an interest has been growing in the direction of theoretical and experimental investigation of gas/shear-thinning fluid stratified flows in pipes (e.g. Xu *et al.* (2007, 2009), Xu (2010), Jia *et al.* (2011) Picchi *et al.* (2014, 2015), Picchi and Poesio (2016a,b)). In those studies, the common approach of one-dimensional two-fluid model (e.g., Heywood and Charles (1979), Bishop and Deshpande (1986)) was used to obtain predictions of the pressure gradient and hold-up. Nevertheless, the two-fluid model requires closures relations for the wall and interfacial shear stresses, which are usually obtained in analogy with the single-phase flow. These may not correctly represent the consequences the interaction between the two-phases, in particular for inclined flows. To account for two-phase interaction, Ullmann *et al.* (2004) proposed the Modified Two-Fluid Model (MTF) closures relations for Newtonian systems, which were derived based on the exact solution for laminar stratified flow between two parallel plates. An attempt to improve the modelling of closure relations for non-Newtonian systems was proposed by Picchi *et al.* (2014), where a pre-integrated model, based on the work of Biberg (2007), was provided for the case of stratified flow of turbulent gas and laminar power-law liquid.

Another aspect to be considered in inclined stratified flow is the issue of multiple solutions for the holdup (and pressure gradient), which can be obtained for specified operational conditions. An exact mapping of multiple hold-up regions have not been provided in the literature for stratified flows involving non-Newtonian fluids, as done by Ullmann *et al.* (2003a,b), Thibault *et al.* (2015) and Goldstein *et al.* (2015) for Newtonian systems. The problems encountered when using the two-fluid model predictions for this aim, which are due to the ambiguity of the appropriate closure relations for the shear stresses, were demonstrated in Picchi and Poesio (2016a) for the case of gas/shear-thinning liquid flow. In fact, the study of the stratified flow in a two-plate geometry can represent a benchmark in the modelling attempts, as it is known that the trends of the main two-phase flow characteristics reflect the corresponding trends in pipe flows.

In this paper, the steady and fully developed solutions for horizontal and inclined stratified flows, where one of the two phases is a non-Newtonian shear-thinning fluid, are presented. Practical configurations for both concurrent and counter-current flows are addressed. In gas/liquid systems the non-Newtonian fluid is the heavy phase (e.g., two-phase flow of crude oil and natural gas), while in liquid/liquid systems the non-Newtonian phase is usually the lighter phase (e.g., waxy oil lubricated by aqueous phase). To address a wide range of applications, this study is not restricted to thin non-Newtonian layers. Our main scope is to provide a methodology for dealing with this family of flows as to provide useful trends and the expected behaviour of the integral variables (e.g. holdup, pressure gradient) for different conditions. As mentioned before, the majority of the modelling activity is carried out by considering the shear-thinning fluid as a power-law fluid. The limitations of this assumption are discussed showing that in some cases it leads to unphysical results due to the unlimited growth of the viscosity at low shear rates (with infinite viscosity at zero shear rate). Therefore, the consideration of the zero-shear-rate viscosity limit cannot be ignored. The case studied is a two-plate geometry, but the conclusions can be extended to pipe-flow modelling via mechanistic models (i.e., Two-Fluid models), which are widely used in engineering applications.

Firstly, an algorithm that is based on a spectral collocation method is proposed for computing the steady and fully developed solutions for the general case of a Carreau fluid. A simplified version of the algorithm valid only for power-law fluid is also provided. Then, the effects of the presence the non-Newtonian shear-thinning fluid layer on two-phase stratified flow characteristics are thoroughly examined and discussed. Concurrent and counter-current inclined systems are investigated, including the mapping of multiple solution boundaries. Aspects relevant to practical applications are addressed, like the trends of the hold-up and frictional pressure gradient variation with the operational conditions, lubrication of a viscous non-Newtonian fluid flow by adding a less viscous Newtonian phase and two-fluid model closures implied by the exact solution. Finally, an algorithm which is based on the two-fluid model is provided to give correct holdup (and pressure gradient) predictions. It includes estimation of the operational conditions where the shear-thinning fluid behaves as a Newtonian fluid, thus allowing the use of the well-known closures relations for Newtonian fluids. The algorithm enables estimation of the expected average shear rate for stratified two-phase flows, which otherwise cannot be a priori obtained without solving the full problem of shear thinning stratified flow.

## 2. Exact solution for laminar stratified flow

*2.1. Problem formulation and rheology model for the non-Newtonian fluid*

The steady and fully developed velocity profiles $u_j(y)$ for two-phase incompressible laminar-laminar stratified flow in a two-plate geometry, see Fig.1, are derived from the solution of the Navier-Stokes equations (in the flow direction, *z*):

$$\frac{\partial \tau_{H,zy}}{\partial y} = \frac{dp}{dz} - \rho_H g \sin\beta, \qquad 0 \le y \le h, \tag{1a}$$

$$\frac{\partial \tau_{L,zy}}{\partial y} = \frac{dp}{dz} - \rho_L g \sin\beta, \qquad h \le y \le H, \tag{1b}$$

with boundary and interface conditions

$$u_H(0) = 0, \qquad u_L(H) = 0, \qquad \tau_{H,zy}(h) = \tau_{L,zy}(h), \qquad u_H(h) = u_L(h), \tag{2}$$

where $dp\ /\ dz$ is the pressure gradient, $\beta$ is the inclination angle to the horizontal, $\rho_{H,L}$ is the phase density, $g$ is the gravity acceleration, $\tau$ is the shear stress and the subscripts (*H, L*) identify the heavy and the light phase, respectively. Steady stratification is considered, where the heavy phase flows at the bottom. Referring to Fig. 1, the inclination angle $\beta$ is always considered positive. Accordingly, in co-current downward inclined flow the flow rates of both phases are positive, while in case of upward inclined flow the flow rates are both negative. In counter-current flows the heavy phase flows downward ($q_H > 0$) and the light phase flows upward ($q_L < 0$).

In this work, one of the phases is assumed a Newtonian fluid, and the other a non-Newtonian time independent shear-thinning fluid with negligible viscoelasticity. Amongst the viscosity models suggested in the literature to model the rheological behaviour of a shear-thinning fluid, we chose to consider the Carreau (1972) model. As in any other rheological model, also in the Carreau model, the model parameters are obtained by curve- fitting to data (available for some range of shear rates), which is an underlying source of error in its predictions. However, this model (see Nouar *et al.,* 2007), has been developed based on the Lodge's molecular network theory, and is capable to model a complex rheology. In particular, it describes well the rheology of the majority of shear-thinning fluids used in applications. The main advantage of the Carreau model is its capability to model the low-shear rate and high shear rate constant viscosity behaviour of a shear-thinning fluid. The constitutive relation of a Carreau fluid for a unidirectional flow in the form of a generalized Newtonian fluid is given by

$$\tau_{j,zy} = \mu_j(\dot{\gamma}_j) \cdot \dot{\gamma}_{j,zy}\,, \qquad \mu(\dot{\gamma}) = \mu_\infty + (\mu_0 - \mu_\infty)\left(1 + (\lambda\dot{\gamma}_j)^2\right)^{\frac{n-1}{2}}, \tag{3}$$

where $\mu_\infty$ is the infinite (shear-rate) viscosity and $\mu_0$ is the zero (shear-rate) viscosity, respectively (i.e., the fluid behaves as a Newtonian fluid at very low and very high shear rates). The index $n$ represents the degree of shear-thinning, while the time constant $\lambda$ indicates the onset of the shear-thinning behaviour (for high values of the time constant the shear-thinning behaviour is shifted to lower shear rates and vice versa). The shear rate for a unidirectional flow is $\dot{\gamma}_{j,zy} = (du_j/dy)$, and the norm of the strain-rate tensor (used in the definition of the generalized viscosity model) is $\dot{\gamma}_j = |\,du_j/dy\,|$.

Despite the power-law model is widely used for shear-thinning fluids, its validity is restricted only to a limited range of shear rates, and predictions resulting from extrapolation beyond that shear rate range (e.g., to low or high shear rates) may provide unphysical results. In fact, the effective viscosity tends to infinite at zero shear rate and to zero at infinite shear rates. Only when $\mu_0$ is much higher than $\mu_\infty$ and $\lambda$ is sufficiently large, the Carreau model reduces to the power-law model for sufficiently high shear-rates, yielding $\mu(\dot{\gamma}) = \kappa_j \dot{\gamma}_j{}^{n-1}$ with $\kappa_j = \mu_0 \lambda^{n-1}$ .

Equations (1a) and (1b) cannot be solved analytically for the general case of a Carreau fluid. An analytical solution can be obtained only for the particular case of a power-law fluid with a positive (or negative) shear rate in the non-Newtonian layer, which requires a priori assumption on the location of the maximum in the velocity profile (e.g., Yu and Dae Han (1973)). In the following section, a numerical approach is proposed to find the exact solution for the general case of stratified flow of a Newtonian/shear-thinning (Carreau) fluid in an inclined channel, without any restrictive hypothesis on the velocity profile. For the sake of brevity, the numerical procedure is presented for the configuration of a heavy non-Newtonian fluid flowing with a lighter Newtonian one (typical to gas-liquid systems). The adaptation for the other case of a lighter non-Newtonian fluid flowing with a heavier Newtonian one (typical to oil-water systems) is rather straightforward.

### *2.2. Problem normalization and solution*

In the following, the laminar dimensionless solution is presented for the case of a heavy non-Newtonian/light Newtonian stratified flow. The normalization of Eqs. (1) is based on single phase flow (i.e., superficial velocity) of the Newtonian phase accordingly to Ullmann *et al.* (2003b) yielding:

$$\frac{\partial}{\partial \tilde{y}}\left\{\tilde{\mu}\left[\tilde{\mu}_0^{\infty}+(1-\tilde{\mu}_0^{\infty})\left(1+\left(\tilde{\lambda}^L\left|\frac{d\tilde{u}_H}{d\tilde{y}}\right|\right)^2\right)^{(n-1)/2}\right]\frac{d\tilde{u}_H}{d\tilde{y}}\right\}=12(\tilde{P}_L-Y_L), \qquad 0\le\tilde{y}\le\tilde{h}, \tag{4a}$$

$$\frac{d^2\tilde{u}_L}{d\tilde{y}^2}=12\tilde{P}_L\,, \qquad \tilde{h}\le\tilde{y}\le 1, \tag{4b}$$

$$\tilde{P}_L=\frac{dp/dz-\rho_L g\sin\beta}{(-dp_f/dz)_{Ls}},\; Y_L=\frac{(\rho_H-\rho_L)g\sin\beta}{(-dp_f/dz)_{Ls}},\; \tilde{\mu}=\frac{\mu_0}{\mu_L},\; \tilde{\mu}_0^{\infty}=\frac{\mu_\infty}{\mu_0},\; \tilde{\lambda}^L=\frac{\lambda|U_{Ls}|}{H}, \tag{4c}$$

$$\tilde{y}=\frac{y}{H},\; \tilde{h}=\frac{h}{H},\; \tilde{u}_j=\frac{u_j}{U_{Ls}},\; U_{js}=q_j/H,\; j=H,L, \tag{4d}$$

where $q_j$ is the volumetric flow rate, $\tilde{u}_j$ is the phase velocity normalized with light phase superficial velocity, $U_{Ls}$, and $(-dp_f\,/\,dz\,)_{Ls}=12\,\mu_L q_L\,/\,H^3$ is the frictional pressure gradient for single phase flow of the light phase ($\mu_L$ is the viscosity of the light Newtonian phase), $\tilde{h}$ is the dimensionless heavy phase height, which in the two-plate geometry coincides with the heavy phase holdup. The inclination parameter and the dimensionless driving force in the light phase are $Y_L$, $\tilde{P}_L$, respectively (the subscript $L$ denotes normalization with respect the light phase). The system characteristic viscosity ratios $\tilde{\mu}$ and $\tilde{\mu}_0^{\infty}$ are constant, while the dimensionless time constant $\tilde{\lambda}^L=\tilde{\lambda}^L(\lambda,U_{Ls},H)$ changes depending on flow conditions, as it depends on the superficial velocity of the light phase. This normalization allows us to provide straightforwardly an interpretation of the results in the framework of two-phase flow and holds both for gas-liquid and liquid-liquid cases. In the following, the solution is calculated using the dimensionless formulation of the equations (Eqs. 4 and 5). The results are provided in a dimensionless form (e.g. holdup and dimensionless pressure gradients trends). In addition, the validity regions of the rheological models are presented on flow maps (in the commonly used coordinates of the fluids' superficial velocities as).

The mass conservation equations of the two phases read:

$$\int_0^{\tilde{h}}\tilde{u}_H(\tilde{y})\,d\tilde{y}=q, \tag{5a}$$

$$\int_{\tilde{h}}^{1}\tilde{u}_L(\tilde{y})d\tilde{y}=1\,, \tag{5b}$$

where $q=q_H\,/\,q_L$ is the flow rate ratio.

As shown by Frigaard and Scherzer (1988) in terms of variational formulation, a solution of the problem is guaranteed for a shear-thinning fluid of a monotone constitutive relationship. In general, the problem is described by eight dimensionless parameters $(Y_L,q,\tilde{\mu},n,\tilde{\lambda}^L,\tilde{\mu}_0^{\infty},\tilde{h},\tilde{P}_L)$, additional three parameters when compared to the Newtonian/Newtonian case, but, for a physical interpretation of the results, we considered the parameters $(Y_L,\tilde{\mu},n,\tilde{\lambda}^L,\tilde{\mu}_0^{\infty})$ fixed, which is the case when the fluids properties, the flow rate of the light phase, the channel height and inclination angle are prescribed. In this case, the solution obtained for $(\tilde{h},\tilde{P}_L)$ vs. $q$ provide their variation with the non-Newtonian fluid flow rate. Note that the parameter $\tilde{\mu}_0^{\infty}$ is relevant only at extremely high shear rates, and for most practical two-phase flow applications (e.g., gas-liquid systems) can be considered to be of a zero value.

#### *2.2.1. Description of algorithm based on collocation points*

From a computational point of view, it is easier to obtain $q$ and $\tilde{P}_L$ in terms of $(\tilde{h},Y_L,\tilde{\mu},n,\tilde{\lambda}^L,\tilde{\mu}_0^{\infty})$. The iterative scheme, which is applied to obtain the solution, is shown in Fig. 2a. In this scheme, the dimensionless pressure $\tilde{P}_L$ is found by solving the mass conservation of the light phase, Eq. (5b), in the form of $F_1(\tilde{P}_L, \tilde{h}) = 0$. Then, the flow rate ratio, $q$ is computed in another loop solving the mass conservation of the heavy phase, Eq. (5a), in the form of $F_2(q, \tilde{P}_L, \tilde{h})$=0 (the integral in Eq. 5a has to be calculated numerically). In these loops, a search for the solution of an implicit algebraic equation is carried out using the Dekker and Brent's algorithm (see Brent (1973)). Note that for a prescribed liquid holdup there is always a single solution for $q$ and $\tilde{P}_L$.

Both iteration loops contain an inner loop, which finds the velocity profiles satisfying the no-slip condition at the interface ($\tilde{u}_{iH} - \tilde{u}_{iL} = 0$) that are then used to find the unknown interfacial shear stress $\tilde{\tau}_i$ ($\tilde{\tau}_i = (d\tilde{u}_L/\tilde{y})|_{\tilde{h}} = \tau_i H/U_{Ls}\mu_L$). In fact, when $\tilde{P}_L$ and $\tilde{h}$ are known, the no-slip condition is a monotonic function of $\tilde{\tau}_i$ (see Taghavi *et al.* (2009) and Alba *et al.* (2013), and the interfacial velocity $\tilde{u}_{i,H,L}$ is only a function of $\tilde{\tau}_i, \tilde{P}_L, \tilde{h}$ . The interfacial velocity of the Newtonian layer is obtained analytically, $\tilde{u}_{iL} = 6\tilde{P}_L\tilde{h}^2 + (\tilde{\tau}_i - 12\tilde{P}_L\tilde{h})\tilde{h} + (-6\tilde{P}_L - \tilde{\tau}_i + 12\tilde{P}_L\tilde{h})$, while $\tilde{u}_{iH}$ of the shear-thinning layer has to be computed by solving numerically Eq. (4a). The latter is determined by imposing the no-slip condition at the wall and an interfacial stress, which is equal to that obtained in the Newtonian layer. Equation (4a) is discretized by spectral methods on Chebyshev collocation points (the discretization is performed following the approach of the Chebyshev first-derivative matrix, see Section 2.4.2 Canuto et al (2006)). The nonlinearity is handled by Broyden's method (see Quaternoni et al. (2007)). Twelve collocation points are sufficient to assure convergence of the steady state solution, upon using $\tilde{u}_{iH}(\tilde{y}) = a \cdot \tilde{u}_{H,Newtonian\,\tilde{\mu}}(\tilde{y})$ , or $\tilde{u}_{iH}(\tilde{y}) = a \cdot \tilde{u}_{H,power-law}(\tilde{y})$ as initial guess for the Broyden's method ($\tilde{u}_{H,Newtonian\,\tilde{\mu}}$ is the Newtonian velocity profile corresponding to the value of the characteristic viscosity ratio, $\tilde{\mu}$, and $\tilde{u}_{H,power-law}$ is the power-law velocity profile corresponding to the viscosity ratio $\tilde{\mu}^L$, see Eq. (A.1) of Appendix A). The parameter $a$ is of the order of 1 and can be tuned to assure convergence.

A more cumbersome procedure is required to solve for $\tilde{h}$ and $\tilde{P}_L$ in terms of $(q, Y_L, \tilde{\mu}, n, \tilde{\lambda}^L, \tilde{\mu}_0^\infty)$, which is described in Fig. 2b. In this case, an additional outer iteration loop is required, where $\tilde{h}$ is computed to satisfy the prescribed $q$ value, i.e., $\Delta q = \left(q - q_{cal}(\tilde{P}_L, \tilde{h})\right) = 0$, and, $\tilde{P}_L$, $q_{cal}$ are obtained from $F_1(\tilde{P}_L, \tilde{h}) = 0$ and $F_2(q_{cal}, \tilde{P}_L, \tilde{h})$=0, respectively. A flow rate ratio $q_{cal}$, which satisfies $\Delta q = 0$ provides the solution for the holdup, and the corresponding $\tilde{P}_L$ are computed using $F_1(\tilde{P}_L, \tilde{h}) = 0$. The convergence is easily guaranteed due to the fact that the flow constrain loops ($F_1$ and $F_2$) have only one solution for a given value of $\tilde{h}$. However, since the solution for the liquid holdup $\tilde{h}$ is not known a priori, a multi-value of $\tilde{h}$ corresponding to the prescribed $q$ can be obtained in concurrent inclined flows, as the outer loop, $\Delta q = 0$ can have multiple solutions. In counter-current flows no-solution for $\tilde{h}$ is obtained for q<0 beyond the flooding point.

A simplified version of the algorithm, which is valid only for a non-Newtonian power-law fluid, is presented in Appendix A. In that case, the solution of the inner loop described above is not needed and the expressions of the velocity profiles and flow rates can be obtained analytically.

*2.3. Pressure gradients*

Once a solution for $\tilde{P}_L$and $\tilde{h}$ has been obtained, the dimensionless frictional and gravitational pressure gradients (the total pressure gradient is the sum of the two contributions) are calculated:

$$\Pi_{fL} = \frac{(-dp_f/dz)}{(-dp_f/dz)_{Ls}} = -\tilde{P}_L + Y_L\tilde{h}\,, \quad \Pi_{gL} = \frac{(dp_g/dz) - (-dp_g/dz)_{Ls}}{(-dp_f/dz)_{Ls}} = -Y_L\tilde{h}\,, \tag{6a}$$

$$\Pi_L = \frac{(-dp\ \ /dz)}{(-dp\ \ /dz)_{Ls}} = \frac{(-dp_f/dz) + (-dp_g/dz)}{(-dp_f/dz)_{Ls} + (-dp_g/dz)_{Ls}} = \frac{\Pi_{fL} - \tilde{h}Y_L - Y_L/(\tilde{\rho}-1)}{1 - Y_L/(\tilde{\rho}-1)}, \tag{6b}$$

where the gravitational pressure gradient is $(dp_g/dz) = [\rho_L + (\rho_H - \rho_L)\tilde{h}]g\sin(\beta)$ , and $\tilde{\rho} = \rho_H/\rho_L$is the density ratio. The above factors represent the 'penalty' in pressure gradient due to the addition of shear-thinning (heavy) phase to the flow of the (light) Newtonian phase. When the lubrication effect is of interest, namely, the potential of pressure gradient reduction achievable by introducing a (light) Newtonian phase to the flow a viscous (heavy) shear-thinning liquid, the frictional pressure gradient normalized by the shear-thinning (heavy) single phase flow can be examined, yielding:

$$\Pi_{fH} = \frac{(-dp_f/dz)}{(-dp_f/dz)_{Hs}} = -\tilde{P}_H + Y_H\tilde{h} \tag{7a}$$

with

$$\tilde{P}_H = \frac{dp/dz - \rho_L g \sin\beta}{\left(-dp_f/dz\right)_{Hs}} \text{ and } Y_H = \frac{(\rho_H - \rho_L) g \sin\beta}{\left(-dp_f/dz\right)_{Hs}} \quad (7b)$$

where $(-dp_f\ /\ dz\ )_{Hs}$ is the exact single-phase frictional pressure gradient of the heavy phase (Carreau fluid).

## 3. Results and discussion: laminar exact solutions

The laminar flow solutions for different practical cases are used for investigating the effect of rheology of the fluid on the stratified flow characteristics, such as hold-up curves, pressure gradient, lubricating effects, and multiple solutions in inclined flows. Our focus is to show the trends that are of interest for applications. The majority of the experimental data available in the literature concerns gas/shear-thinning fluid stratified flow (see Table 1, where a summary of published experimental work is reported). Solutions of water and carboxymethyl cellulose (CMC) polymer are widely used to create shear-thinning test fluids. However, those studies considered a power-law rheological model, while the zero-shear viscosity limit was ignored.

**Table 1.** Summary of published experimental work on gas/shear-thinning fluid stratified pipe flows.

| Solution | polymer | $n$ (-) | $\kappa$ (Pa s$^n$ ) | $D$ (m) | $\beta$ (deg)* |
|---|---|---|---|---|---|
| Bishop and Deshpande (1986) | 74H SCMC | 0.68 - 0.85 | 0.024 – 0.139 | 0.025 – 0.052 | 0 |
| Chhabra *et al.* (1983) | kaoline | 0.103 | 19 - 48.5 | 0.207 | 0 |
| Xu *et al.* (2007) | CMC | 0.615- 0.798 | 0.972 – 0.089 | 0.02– 0.06 | 0, +5, +15,+30 |
| Xu *et al.* (2009) | CMC | 0.53 – 0.952 | 2.434 – 0.034 | 0.05 | 0 |
| Jia *et al.* (2011) | CMC, kaoline | 0.17 – 0.952 | 4.25 – 0.034 | 0.05 | 0 |
| Picchi *et al.* (2015) | SCMC | 0.757 – 0.942 | 0.264 – 0.007 | 0.0228 | 0, +5 |

*Positive $\beta$ refers to downward inclined flow.

In the present analysis, we refer to real shear-thinning fluids whose full rheological curve is available in the literature (e.g., Sousa *et al.* (2005)). The rheological data obtained by Sousa *et al.* (2005) for CMC solution are provided in Table 2. These data include the behaviour of the effective viscosity also at low shear rates (where the fluid behaves as a Newtonian one). We chose to fit the data with the Carreau model, which provides a good representation for the rheological behaviour of many polymer solutions and a negligible infinity viscosity is assumed. The results shown in Table 2 clearly indicate that the rheological parameters of the Carreau model $(\mu_0, \mu_\infty, \lambda, n)$ are not independent: with increasing, both the zero-viscosity and the time constant increase, while the flow behaviour index decreases. Other solutions (e.g xhantam gum) can produce various combinations of the Carreau fluid parameters, but we will focus on high viscous water-CMC solutions, which are commonly used in experiments (note that the slightly viscoelasticity of the CMC08, i.e., 0.8% weight CMC, solution is neglected, considering that the fully-developed flow is obtained after the relaxation time has elapsed). A more complex fluid, a shear-thinning stabilized oil-in water emulsion, from Partal *et al.* (1997) is also considered. The idea is to examine the stratified flow characteristics for two-phase systems used in applications, where the non-Newtonian fluid is an effectively a viscous shear thinning-fluid.

**Table 2.** Carreau viscosity model and power-law parameters for different CMC solutions by Sousa *et al.* (2005) and a stabilized oil-in-water emulsion by Partal *et al.* (1997).

| Solution | $\mu_0$ (Pa s) | $\mu_\infty$ (Pa s) | $\lambda$ (s) | $n$ (-) | $\kappa$ (Pa $s^n$) |
|---|---|---|---|---|---|
| CMC01 | 0.0091 | 0.0 | 0.0485 | 0.8975 | 0.0124 |
| CMC03 | 0.0484 | 0.0 | 0.0902 | 0.7556 | 0.0871 |
| CMC05 | 0.2043 | 0.0 | 0.1950 | 0.6140 | 0.3840 |
| CMC08 | 0.9831 | 0.0 | 0.4639 | 0.5116 | 1.4306 |
| 3%SE 60% O/W Emulsion (35°) | 0.7230 | 0.0181 | 0.4583 | 0.4341 | - |

*The data by Sousa et al. (2005) were fitted using the Carreau model instead of the Carreau-Yasuda model.

### *3.1. Horizontal gas/shear-thinning fluid stratified flow*

In this Section, we will examine how the rheology of the shear-thinning liquid affects the flow characteristics of gas/liquid in the case of horizontal stratified flows.

Figure 3(a) shows the holdup curve ($\equiv\tilde{h}$ in a two-plate geometry) obtained for horizontal flow of air and the CMC05 solution as a function of the flow rate ratio $q$ (a change of the flow rate ratio corresponds to a change in the liquid, heavy phase, superficial velocity). In Figure 3(b) the trend of the shear rate at the (lower) wall, at the interface and the average shear rate in the liquid layer ($\bar{\dot{\gamma}} = 1/h \cdot \int_0^h \dot{\gamma} dy$) are presented, while the viscosity curve of the CMC05 solution is shown in Fig. 4. At a fixed holdup, the shear rate varies along the non-Newtonian layer (i.e., as a function of $\tilde{y}$), but the general behaviour can be studied by examining the average shear rate, or the shear rate values at the wall and at the interface. Inspection of Fig. 3 indicates that the studied shear-thinning liquid behaves practically like a Newtonian fluid over a wide range of holdups: at low holdups, the wall (and the average) shear rate in the liquid corresponds to the zero-viscosity ( $\mu_0$ ) region of the viscosity curve, Fig. 4, resulting in a Newtonian behaviour. Only at sufficiently high holdups (hence shear rates) the shear-thinning effect on the holdup becomes evident. Note that for the CMC solutions, the highest shear rates studied are still far from the extremely high shear rate range, where the fluid is expected to reach the infinity viscosity limit (assumed zero in the rheological model considered here). The effect of the dimensionless time constant on the holdup curve is shown in Fig.5. As expected, at low flow rate ratios, when the liquid behaves as a Newtonian fluid, the holdup curves converge to a single curve. However, a different trend is observed in the shear-thinning region, whereby increasing the light (Newtonian) phase superficial velocity results in a lower holdup for the same flow rate ratio.

Figure 6 shows a comparison of the obtained holdup curve with those predicted by assuming Newtonian behaviour with viscosity $\mu_0$ (denoted as zero-viscosity), and by the power-law model. Such a comparison enables evaluating the validity of using those asymptotic simpler rheological models, which are widely used in the literature for the modelling of the integral two-phase flow characteristics (e.g., holdup and pressure gradient). The transition between the two asymptotic behaviours is also highlighted in Fig. 6. As shown in the figure, even though the fluid has a complex rheology (e.g., Carreau viscosity model), depending on the operational conditions, the two-phase stratified flow may exhibits a Newtonian behaviour of the liquid phase (i.e., viscous effect corresponding to the zero-viscosity), or a power-law behaviour (i.e., the shear-thinning effect due to the decreasing of the effective viscosity at high shear rates). There is only a limited range of operational conditions where the transition between the two asymptotic behaviours takes place, where the fluid should be described by the Carreau model.

Indeed, the modelling of two-phase stratified flows has to take into account this feature. If the system is operating at low shear rate conditions, considering the power-law model instead of the Carreau model, gives a dramatic over-prediction of the effective viscosity (it tends to infinity) and the resulting predicted plug-like velocity profile is not realistic. On the other hand, if the system is operating in the shear-thinning region, the power-law is a valid model to correctly represent the effects of the viscosity decrease with the shear rate on the holdup. The intersection between the two asymptotes in Fig. 6 corresponds to the conditions where both the power-law and the Newtonian models fail to represent the real (Carreau fluid) behaviour. This is demonstrated in Fig. 7, where the velocity profiles corresponding to those operational conditions are plotted. As shown, the exact velocity profile is very different compared to the asymptotic ones, indicating that an incorrect use of the rheological model can result in large errors in the predicted flow characteristics. Examining the corresponding effective viscosity profile (Fig. 7(b)) indicates that the power-law model fails at the interface, where the viscosity is highly over-predicted and locally the fluid is practically Newtonian. On the other hand, the Newtonian model overestimates the viscosity at the wall. This reflects a trend in laminar gas-liquid systems, where the shear rate at the phases interface is much smaller compared to the liquid shear rate at the wall, and the interface region exhibits a Newtonian behaviour for a large variety of operational conditions (see Fig 3(b)). Thus, the zero-shear-rate viscosity behaviour cannot be ignored in stratified flow. Unfortunately, differently from single-phase flow, where it is possible to estimate a priori the order of magnitude of the expected shear rates, in two-phase flows, the shear rates are very much dependent on the holdup and the resulting interaction between the phases. Consequently, it is difficult to a priori determine the range of shear rates in the non-Newtonian layer. To tackle this difficulty, a methodology for estimating the range of operational conditions where either the Newtonian or the power-law asymptotic models are applicable, is proposed in Section 4.2 below.

The behaviour of the two-phase flow system becomes clearer when the results are depicted in the phases' superficial flow rate coordinates, which are commonly used in flow pattern maps. In Fig. 8, the Newtonian and the shear-thinning region are mapped. In this figure, the lower (upper) boundary of the transition region marks the region below (above) which the discrepancy between the holdup predictions obtained by the Newtonian (power-law) fluid model are less than 0.25%, respectively (when compared to the Carreau model). Note that in this map the results shown in Fig. 3(a) (i.e., a constant $U_{Ls}$) correspond to a vertical line of Fig. 8. This figure shows that for a wide range of operational conditions (low superficial liquid velocities) the two-phase flow behaves practically as stratified flow of two Newtonian fluids. Only at higher liquid (and gas) superficial velocity, due to the increased shear rates, the shear thinning effect emerges and the effective viscosity decreases in accordance to Fig. 4. The iso-holdup lines are straight lines in the Newtonian region, and start diverging upon approaching the shear-thinning region. In any case, due to the high value of the zero-viscosity, the hold-up is very high (> 0.9) for a wide range of operational conditions. The so-called 'asymptotic boundary' between the Newtonian and shear-thinning region represents the locus of the intersection of the asymptotes shown in Fig. 6. This boundary follows more or less the contours of the effective viscosity when based on the average shear rate, and corresponds to $\mu_{H,eff}/\mu_o \approx 0.9$. This information will be useful for validating the criterion presented in Section 4.2.

Figure 9 shows the frictional (i.e., the total for $Y_L = 0$) pressure gradient for a gas/shear-thinning liquid stratified flow corresponding to Figs. 3(a). The trends discussed with respect to the holdup curve are observed also in the pressure gradient. However, the errors introduced by using an incorrect viscosity model (e.g., using the simpler power-law model for modeling the shear thinning behavior and ignoring the zero-viscosity limit) can reach up to 100% in the pressure gradient predictions. A reasonable simplification that can be adopted is to consider only the asymptotic rheological models, where the switch between the two is taken at the point of intersection of the two asymptotic curves. As shown in Fig 10 the errors in the holdup and pressure gradient predictions (with reference to the exact (Carreau) solution), when evaluated along the asymptotic boundary of Fig. 8, are still acceptable.

*3.1.1. Effects of liquid rheology and channel size*

In general, a different geometry influences the shear-rate distribution within shear-thinning fluid. This is demonstrated in Fig. 11(a) where the effect of the channel size is shown. Increasing $H$ (while maintaining the same superficial velocities) reduces the shear rate at the wall, and consequently the Newtonian region expands to higher liquid superficial velocities. Likewise, for the case of a horizontal mini-channel, $H = 0.002$ m, the Newtonian region shrinks to much lower superficial liquid velocities. Similar trends are obtained when considering the other fluids included in Table 2 (not shown for the sake of brevity).

Figure 11(b) presents the asymptotic boundary and the rheological regions for different shear-thinning solutions given in Table 2 for horizontal flow. With increasing the polymer concentration in the CMC solution, the Newtonian region moves to lower liquid superficial velocities and the transition (Carreau) region becomes wider. Also, due to the increased zero-viscosity value, the holdup obtained for the same flow conditions is higher. Although the rheological parameters that characterize the non-Newtonian fluid ($\mu_0, \lambda, n$)are not independent, the qualitative trends affected by their variation can be deduced from Fig. 11(b).

*3.2. Downward inclined gas/shear-thinning fluid stratified flow*

The majority of the experiments of gas/shear-thinning stratified flows were performed in downward inclined pipes, see Table 1. Fig. 12(a) shows the hold-up curve for a downward inclined channel. The shear-thinning effect is evident over the majority of conditions, i.e. the liquid behaves practically as a power-law fluid, except for very low liquid superficial velocities and hold-ups, where it exhibits a Newtonian behavior. In addition, the transitional region, where the Carreau model should be used to represent the liquid behavior, corresponds to rather low flow rate ratios. Fig. 12(b) shows that apparent power-law rheological behavior is actually determined by the shear rate at the wall, while the shear rate at the interface corresponds mainly to the Newtonian region. Examining Fig. 12(a) indicates that the apparent Newtonian region corresponds to very low liquid superficial velocities that are commonly not encountered in experiments. The dominancy of gravity in the range of the flow conditions of practical interest is manifested in the holdup curve: the slight downward inclination changes dramatically the holdup as compared to a horizontal channel, while the horizontal and inclined systems holdup curves converge only at very high values of the flow rate ratio, where gravity effects become negligible. As shown in Fig. 13(a) the locus of the iso-hold-up lines is mostly independent of the light phase (air) superficial velocity, and the approach to the shear-dominated flow region becomes noticeable only at extremely high hold-ups and high air superficial velocities. In fact, the shear-thinning nature of the liquid enhances the effect of gravity, which accelerates the flow and reduces the liquid holdup.

The window of operational conditions were experiments were conducted in downward inclined flows (see Table 2) is also indicated in Fig. 13, where two different shear-thinning fluid are considered. As implied by the figure, in those experiments the shear-thinning liquid can be modelled as a power-law fluid to obtain holdup (and pressure drop) predictions. Since, the 1D Two-Fluid model is widely considered for that purpose, the rheological region map can serve as a guide for the selection of appropriate closure relations when such mechanistic models are applied for predicting the two-phase characteristics. Although the automatic choice of the power-law based closure relations when a shear thinning fluid is involved is not always correct (e.g., horizontal flow, see Section 3.1), it represents a valid approximation for downward inclined flow in the flow rates range of interest.

Due to the competition between gravity and shear forces, multiple solutions for the same flow conditions can be obtained in inclined flows (as shown in Fig. 12 (a)). The boundaries of triple-solution (3-s) region in the downward inclined system for the conditions of Fig. 13 are indicated in the figure. As shown, this region corresponds to low gas and high liquid superficial velocities. When operating with shear-thinning liquids gravity effects are enhanced, and consequently the 3-s region expands to higher gas (and liquid) flow rates, as compared to 3-s region obtained when a Newtonian liquid is assumed (with $\mu_H = \mu_0$). This is consistent with the reported effects of decreasing the (Newtonian) liquid viscosity on the 3-s region boundaries (Ullmann *et al.* (2003a)). Nevertheless, the 3-s region is out of the experiments operational window. Indeed, multiple holdups were not encountered in experiments of Xu *et al.* (2007) and Picchi *et al.* (2015) in downward inclined flows. Obviously, as the 3-s region corresponds to high holdup of the liquid, small perturbations at the interface may result in a transition to slug flow.

An example of velocity profiles corresponding to the multiple solutions is given in Fig. 14. As shown, the liquid velocity profiles practically coincide with those predicted by the (simpler) power-law rheological model. Hence, the latter can provide a reliable prediction of the triple solution region in downward inclined flows and can be applied in this region in the framework of two-fluid models (see Picchi and Poesio (2016)). As shown in Fig. 14, the liquid velocity profiles (corresponding to the 3 holdup solutions) show the typical trend of a shear-thinning fluid, with a tendency to have a plug-like region near the phases interface, and backflow (upflow) of the light phase near the upper wall is present.

Because of gravity, the anticipation of the validity range of the liquid rheological model in inclined channels is rather not intuitive. For example, the effect of the channel height $H$ in downward inclined flows has an opposite effect on the apparent liquid rheological behavior compared to that in horizontal flows. In a larger channel, the transition from Newtonian to the shear thinning behavior is anticipated at lower flow rate ratio (i.e., at lower liquid superficial velocity in the flow pattern map). This is due to the effect of the gravitational body force, which accelerates the liquid. In fact, by increasing $H$, the flow becomes more gravity dominated (similarly to increasing the inclination angle, since $Y_L \propto H^2 \sin(\beta)$).

*3.3. Upward inclined gas/shear-thinning fluid stratified flow*

Figure 14 shows the holdup curves for the case of upward inclined flow of gas/shear-thinning (CMC05) liquid for two different inclinations. In the first case, a very shallow inclination affects dramatic changes in the two-phase flow characteristics.

As shown in Fig. 15 (a), at very low flow rate ratios, a triple solution for the holdup is obtained for the same $q$ and countercurrent flow ($q < 0$) with two solutions is also feasible. In terms of the rheological behavior, in the entire region of low flow rate ratios, the non-Newtonian fluid behaves like a zero-viscosity Newtonian fluid, while the shear-thinning effect becomes dominant only at high $q$ and high hold-up. Practically, the region of multiple solutions (3-s) is obtained at very low liquid superficial velocities. When the polymer concentration is increased (i.e., the zero-viscosity value increases), the 3-s region moves to even lower liquid superficial velocities (see Fig. 16). This trend is consistent with the results of Ullmann *et al.* (2003a) since the shear-thinning fluid behaves practically as a Newtonian one under those conditions.

The presence of multiple solutions in upward flows is associated with backflow of the heavy shear-thinning phase. An example of velocity profiles of multiple solutions is given in Fig. 17(a). Therefore, since the 3-s region is within the Newtonian region, the two-fluid model with the Modified Two Fluid (MTF) closures relations (Ullmann *et al.* (2004)) can be suggested as a reliable tool for predicting the two-phase flow characteristics in this region. Upon increasing the inclination angle, the 3-s region moves to higher gas and liquid superficial velocities, but in that case, the turbulence of the gas phase should be considered in the modelling (see Section 4.3). In fact, the location of the 3-s region more or less demarks the gas velocity where stratified flow can be obtained in upward inclined systems. For lower gas superficial velocities, where only a single high-holdup solution is obtained, interfacial disturbances lead to the establishment of slug flow. Indeed, the latter is known to be the dominant flow pattern in upward inclined flows.

The flow characteristics in the counter-current region are demonstrated with reference to Fig. 15(c). If the inclination parameter is sufficiently high, the countercurrent region tends to the power-law asymptote (gravity accelerates the heavy non-Newtonian phase), except at conditions associated with very low $|q|$, which corresponds to a Newtonian behavior (see the shear rates trends of Fig. 15(d)). An example of velocity profile in the countercurrent region is given in Fig. 17(b). As can be seen, even though the hold-up predicted by the power-law asymptote is practically the same as that obtained by the Carreau model, the velocity profiles do not match (in particular for the high holdup solution). Generally, for the modeling of countercurrent flows it is recommended to employ the Carreau rheological model to avoid unnecessary errors in the predictions.

The above conclusion applies also for the modelling of slug flow in upward inclined systems. Picchi *et al.* (2015) observed backflow of the liquid in the film zone of the slug flow in upward inclined gas/shear-thinning fluid system, which corresponds to local counter current flow. In fact, as the system becomes more gravity dominated due to the shear-thinning effect, the tendency of the liquid (i.e., heavy phase) to flow back in upward inclined systems is more pronounced (see Fig. 17(b)) compared to Newtonian fluids, and the flow rate range where counter current flow is feasible expands.

### *3.4. Liquid/liquid Newtonian/shear-thinning fluid stratified flow*

The case of liquid/liquid systems is also briefly addressed for the sake of a complete picture. To this aim, the CMC solutions studied so far cannot be considered, as the polymer is soluble in water and probably in oil. A typical case of liquid/liquid system is the flow of a very viscous shear-thinning emulsion (or waxy oil) lubricated by water. In this case, the Newtonian phase (e.g water) is the heavy one and the shear-thinning fluid (emulsion) is the light one. Rheological data for a shear-thinning emulsion by Partal *et al.* (1997) are presented in Table 2 (and Fig. 18 below). It is important to highlight that for shear-thinning emulsion it is not possible to neglect the infinity viscosity $\mu_\infty$, since at high shear rates that are encountered in the flow, the shear thinning effect is no longer valid (see Fig 18). The solution algorithm outlined in section 2.2.1 can be easily rearranged for the light non-Newtonian fluid configuration. The in situ volume fraction of the (light) shear-thinning fluid corresponds to $1-\tilde{h}$ (which is denoted below as the emulsion holdup).

The holdup curve is presented in Fig. 19 (l.h.s. figure) along with the trends of the shear rates (r.h.s. figure, where the critical values which mark the transition from/to the zero and infinity asymptotic behaviour are indicated). It is clear that the shear thinning effect emerges only at intermediate $1/q$ values, when a change in the slope of the holdup curve is noticeable and the average shear rates are in the range of shear-thinning behaviour. This is even more evident in Fig. 20, where the effective viscosity curves computed based on the average shear rates of Fig. 19(b) are plotted. The effective viscosity based on the average shear rate (black line) starts decreasing when the shear effect is evident also in the holdup curve, but the rheological behaviour at the interface is mostly Newtonian. Only when the holdup of the non-Newtonian layer is extremely high, the shear-thinning effect is evident also at the interface. The effect of the dimensionless time constant on the holdup curves is demonstrated in Fig. 21, where the holdup curves converge only when the shear-thinning effect is negligible, and the non-Newtonian liquid behaves as a Newtonian one. As shown, increased time constant $\lambda^H = \lambda\,|U_{Hs}|/H$ (i.e., higher heavy (Newtonian) phase flow rate, or smaller channel size), results in enhancement of the shear-thinning behaviour. For example, considering the same test fluids and operational conditions, a more pronounced shear thinning behaviour should be considered upon downscaling results obtained in conventional size channel to mini and micro-channels.

The mapping of the rheological regions, which is based on the holdup predictions compared to Carreau model results, is shown Fig. 22. Differently from gas/liquid systems, a transition to the $\mu_\infty$-region appears. This mapping shows the complexity of studying the stratified two-phase flow characteristics in a wide range of flow rates, since there are conditions where the shear-thinning effect is practically absent (in terms of holdup predictions) and others where it is dominant.

Another aspect of interest for liquid/liquid systems is the issue of whether or not a lubrication effect is achievable by adding a less viscous (Newtonian) phase to the flow of a more viscous shear-thinning liquid. In horizontal flows, the frictional pressure gradient coincides with the total pressure gradient, and conditions with $\Pi_{fL} < 1$ indicate a lubrication effect. To this aim, the frictional pressure gradient should be normalized with the single-phase flow of the more viscous shear-thinning phase. For a Carreau fluid, $(-dp_f\,/\,dz\,)_{Ls}$ should be computed numerically, whereby the velocity profile is found iteratively by discretizing the single-phase momentum equation on Chebyshev collocation points (similarly to the inner loop described in Section 2.2) and satisfying the flow rate constrain for the non-Newtonian fluid. For the sake of physical interpretation of the lubrication effect, the flow rate of the non-Newtonian phase is maintained constant, while the consequence changing the flow rate of the lubricating (Newtonian) phase is examined. To this aim the calculation of $\Pi_{fL}$ can be carried out using the solution algorithm that considers normalization with respect to the Newtonian phase (which yields $\Pi_{fH}$) and adjusting the results point-by-point to obtain $\Pi_{fL}$.

Fig. 23 demonstrates the lubricating effect ($\Pi_{fL} < 1$) upon adding water to the flow of a light and more viscous non-Newtonian emulsion: $\Pi_{fL}$ decreases until reaching a minimum and for a wide range of operational conditions a lubricating effect due to the addition of water is observed. In order to show the shear-thinning effect on the lubrication, the curve corresponding to a Newtonian fluid with zero-shear-rate viscosity is also plotted. As expected, a larger pressure reduction is achievable considering a fluid with a higher viscosity (the minimum of the $\Pi_{fL}$ curve is lower), but the lubrication effect is confined to a smaller range of flow ratios (i.e., lubricant flow rates).

In addition, since the flow rate is a multivalued function of the pressure drop, the variation of the pressure drop with the flow rate of the lubricating phase (Fig. 23) implies that the system may be susceptible to Ledinegg instability (Ledinegg, 1938). In fact, the pressure drop curve has a negative slope for lubricant flow rates ranging from zero to the value associated with the minimum pressure drop. Therefore, if a pump of a constant pressure characteristic (or with a larger slope than the system pressure gradient) is used to introduce the (Newtonian) lubricant into the pipe, Ledinegg instability may occurs. In this range of negative slope of $\Pi_{fL}$, any small negative disturbance in the flow rate will result in zero flow rate of the lubricant. On the other hand, any small positive disturbance will result in a system pressure drop smaller than that supplied by the pump, and consequently a spontaneous shift to a much higher flow rate (and higher lubricant holdup) that corresponds to the same pressure drop. Further discussion on the Ledinegg instability in stratified flows can be found in Goldstein *et. al.* (2015).

## 4. Results and discussion: practical implications for the 1D two-fluid model modelling

### *4.1. Velocity shapes factor for gas/shear-thinning fluid stratified flow*

The exact solution of the velocity profiles can be used to calculate the velocity profile shape factors. These should be introduced in the phase's inertia terms of the well-known one-dimensional two-fluid model. The shape factors are relevant in all the calculations that involve the inertia terms, like in simulation of transient or undeveloped flow conditions, or when the TF model is used for stability analysis (e.g., Brauner and Moalem Maron (1992), Holmas *et al.* (2008), Picchi *et al.* (2014)). However, the modelling efforts to predict the shape factor values have been rather poor, in particular, for the case of non-Newtonian fluids. Since the exact 2-D velocity profiles are needed to compute the exact shape factors, they are commonly set to a unity value (plug flow assumption).

The velocity shape factors are defined as:

$$\gamma_L = \frac{1}{(1-\tilde{h})\,\tilde{U}_L^2}\int_{\tilde{h}}^{1} \tilde{u}_L^2(\tilde{y})d\tilde{y}\,, \qquad \gamma_H = \frac{1}{\tilde{h}\,\tilde{U}_H^2}\int_0^{\tilde{h}} \tilde{u}_H^2(\tilde{y})d\tilde{y}\,, \tag{8}$$

where $\tilde{U}_{L,H}$ is the dimensionless phase average velocity. The expressions of the shape factors can be obtained analytically only for laminar flow of Newtonian fluids and they require numerical integration in case of non-Newtonian fluids.

Figure 24 shows the shape factors $\gamma_{L,H}$ values as a function of $\tilde{h}$ obtained for horizontal gas/non-Newtonian shear-thinning fluid. In case of a Newtonian liquid, the value of $\gamma_H$ approaches 4/3 (corresponding to a linear velocity profile) in the limit of a vanishing heavy phase ($\tilde{h} \to 0$). With increasing $\tilde{h}$, it decreases to a minimum value of $\gamma_H < 1.2$ ($\gamma_H = 1.2$ corresponds to a semi-parabolic velocity profile with the maximal velocity at the interface, which for the case shown is close to $\tilde{h}$=1), and finally the single-phase limit of 1.2 is reached (corresponding parabolic velocity profile). The shape factor of the light phase is approximately constant for a wide range of holdups, see Fig 24 (b).

Considering a shear-thinning fluid, the trends of the shape factors changes dramatically. Consistent with the discussion of results obtained for the holdup, Fig 3(a), $\gamma_H$ follows the Newtonian asymptote at low holdups and the power-law behaviour at high holdups. In between, a transition between the two asymptotic behaviours takes place. It is evident that, when the shear thinning effect is predominant (i.e., shear-thinning region), the shape factor of the heavy phase attains lower values compared to Newtonian fluid, due to the tendency to the plug-like shape of the velocity profile. It is to be emphasized that, for power-law fluid the single-phase limit ($\tilde{h} \to 1$) is a function of the fluid behaviour index:

$$\gamma_{H,\tilde{h}\to 1} = \frac{2(2n+1)}{3n+2}, \tag{9}$$

where ($\gamma_{H,n\to 0} = 1$ and $\gamma_{H,n\to\infty} = 4/3$). However, correct modelling of the shape factor should consider the zero-shear-rate viscosity limit, since assuming only a power-law fluid is not sufficient to describe the real trend. The $\gamma_{L,H}$ values for a downward inclined flow are also presented in Fig 24(a,b), where, in accordance with Fig. 12, the majority of flow conditions are in the shear-thinning region and $\gamma_H$ attains lower values (close to 1.1) compared to horizontal flow. The variation of the shape factors in upward inclined flows, Fig. 25, is much more dramatic, in particular for $q \to 0$ when the counter current region is approached. The middle and upper solutions obtained for $q = 0$ correspond to circulating flow ($\tilde{h} \neq 0$, but $U_{Hs}$=0 ), whereby $\gamma_H \to \infty$ for $q \to 0$. This behaviour is obviously common to Newtonian and non-Newtonian rheology of heavy phase, when the inclination parameter is sufficiently high to enable a counter current flow region, and thus a non-zero solution for the holdup at $q = 0$. An adequate modelling of the shape factors is necessary to correctly represent the heavy phase inertia, in particular for the conditions close to the flow reversal, where the $\gamma_H$ values are far from 1.

*4.2. TF model predictions and algorithm based on the TF model to predict the presence of Newtonian behaviour*

The majority of the published works consider the shear-thinning fluid using the power-law model, which erroneously predicts a non-physical infinity effective viscosity at low shear rate conditions. From the practical point of view, it is of interest to a-priori predict if the zero-viscosity behaviour arises for a given operational conditions in order to avoid large mistakes in the predictions of flow characteristics. As demonstrated and discussed above, due to the dependence on the holdup (which in turn depends on the other system parameters), it difficult to a priori estimate the shear rate range in the non-Newtonian fluid. However, a procedure, which is based on the two-fluid (TF) model, is suggested here to provide an indication if, for a specific operational conditions, the power-law model can be considered as a good approximation (without working out the exact solution of the Carreau model). Note that although the shear rate in a shear-thinning fluid changes locally as a function of $\tilde{y}$, it is reasonable to assume that the flow is controlled by the shear stresses (and shear rates) at the wall and at the interface, whereby the average shear rate represents a reasonable gauge.

The proposed algorithm is the following. For given flow conditions ($U_{Hs}, U_{Ls}$):

1. The zero-viscosity Newtonian solution for the holdup is computed using the two-fluid model with the MTF closure relations. The MTF model gives the exact solutions for the case of two-plate geometry, and practically converge to the exact laminar solutions also for a pipe geometry, see Ullmann *et al.* (2004).
2. The "average" shear rate is estimated based on the heavy phase average velocity $U_H$ and the liquid height $h$ (predicted by step 1), whereby $\dot{\gamma}_{av.} = 2U_H/h$ . This estimation assumes a linear velocity profile. This assumption is reasonable for horizontal and downward inclined gas-liquid stratified flows, or for liquid/liquid stratified flow systems, when the effective viscosity of the shear-thinning liquid is much larger compared to the Newtonian phase.
3. If $\dot{\gamma}_{av.} < \dot{\gamma}_{CRIT.}$ ($\dot{\gamma}_{CRIT.}$ is the shear rate that corresponds to the intersection of the zero-viscosity and the power-law asymptotes in the viscosity curve, see Fig. 4(a)), the TF model with the MTF closure relations is the best predictive tool to be used. Otherwise, switch the TF closures relations for a power-law fluid (TF SP-PL), see Appendix C.
4. Verify that the "average" shear rate based on the phase velocity and the liquid height computed by the TF SP-PL exceeds $\dot{\gamma}_{CRIT}$

Note that no theory-based closure relations that account for the two-phase interactions are available for gas/shear-thinning liquid systems. The common practice is to adopt the single-phase based closure relations, where the phase Reynolds number is based on an adjusted hydraulic diameter. For gas/liquid systems, the practice of considering the interface as a free surface with respect to the liquid flow, and as a wall with respect to the gas flow, provides a good estimate of the friction factors. For liquid/liquid systems the hydraulic diameter is defined based on the phases' relative velocities, see Appendix C. The validation of this criterion is shown in Fig. 8 for horizontal flow and Fig. 13 for downward inclined flows, where the boundary which delimits the zero-viscosity Newtonian region predicted by the criterion is indicated. The results show that the boundary obtained by applying the above algorithm is close to the (asymptotic) boundary of the exact solution. Figure 18 shows the results of applying the algorithm to a liquid/liquid system. As shown in the figure, the presence of the zero-viscosity Newtonian regions is predicted by the algorithm with a satisfactory agreement compared to the exact mapping and, therefore, the regions of operational conditions where the emulsion behaves as a Newtonian liquid can a priori be identified.

With the above algorithm, the two-fluid model is a valid tool to predict holdup and pressure gradient for gas/shear-thinning fluid flows (horizontal and downward inclined, when back flow at the wall in the heavy phase is not present). The switch between the MTF closures relations (applicable in the Newtonian region) to the SP-PL closures relations is done following the proposed criterion and, as is shown in Fig. 26, the results are very satisfactory. In the Newtonian region, the MTF reproduces exactly the exact solutions, and in the power-law region the single-phase based closure relations provide predictions that are very close to the exact solution. Obviously, in the Carreau region the small deviations in predictions are unavoidable.

Although the method is tested here for the two-plate geometry, it is applicable also to pipe flows. Accordingly, this algorithm can provide a useful tool for estimating the Newtonian region in order to avoid mistakes in the modelling and in the holdup and pressure drop predictions when the power-law based models and empirical correlations are considered for representing the shear-thinning behaviour of the liquid phase.

### *4.3. Effect of turbulence in the gas layer on gas/liquid stratified flows*

In many practical applications (e.g. in large channels/pipe and for high superficial gas velocity) the gas phase is turbulent, and, therefore, it is of interest to show the qualitative effect of the gas turbulence on stratified flow characteristics. For high gas flow rates Tollmien-Schlichting instabilities that are triggered in the bulk of the gas phase lead to laminar/turbulent transition in the gas (see Barmak *et al.,* 2016). In the present work, the issue of the stratified flow stability is not addressed, but it can be assumed that turbulent gas flow should be considered when the gas superficial Reynolds number exceeds a threshold value (e.g., $Re_{Ls} = (\rho_L U_{Ls} H/\mu_L) > 5772$). For those conditions, the momentum equation in the laminar shear-thinning liquid can be solved, as shown in Appendix D, considering a turbulent gas flow. Figure 27 presents the holdup curves for horizontal flow obtained for both laminar and turbulent gas at the same gas superficial velocity. As expected, upon including the effect of the gas turbulence, the transition from the Newtonian to the power-law behaviour is at lower flow rates of the heavy (shear- thinning) liquid. Also, for the same superficial liquid velocity, the hold-up of the turbulent solution is lower, since the interfacial shear stress exerted by the turbulent gas-is higher.

Thus, for applications where a non-Newtonian shear-thinning layer is sheared by a turbulent gas, the validity region of the power-law model is extended, while the zero-viscosity Newtonian behaviour is shifted to lower liquid superficial velocities.

## 5. Conclusion

In most of modelling works which considered two-phase flow in the presence of shear-thinning fluid, a power law rheological model is adopted for the non-Newtonian phase. In this study, we challenge this assumption. To this aim, a theoretical investigation of Newtonian/non-Newtonian shear-thinning fluid stratified flow in two-plate geometry for horizontal and inclined flows is carried out. The exact solution is calculated considering the general Carreau viscosity model for the shear-thinning fluids. An iterative algorithm is presented and described in details to find the exact laminar solution for the general case of a Newtonian/Carreau fluid stratified flow. All the required computational details are provided in the paper. A series of practical cases are studied to show the two-phase flow characteristics of this family of systems.

The results of the study highlight the fact that even if the liquid is characterized a shear-thinning fluid, in some operational conditions associated with low shear rates the Newtonian behaviour must be considered for a correct prediction of the of the flow characteristics. The results are discussed in terms of holdup-curves and dimensionless pressure gradient, showing the effect of the rheology of shear-thinning liquids on the two-phase flow characteristics. Depending on the operational conditions, the two-phase stratified flow can present a viscous effect (due to the zero-shear-rate viscosity limit), or a shear-thinning behaviour at sufficiently high shear rates, while in a limited range of operational conditions the transition between the two asymptotic behaviour takes place. The validity regions of those rheological models are shown on flow maps. The effects of the inclination and the rheology are discussed for gas/liquid systems, including the issue of multiple holdup solutions in inclined flows, referring to some real shear-thinning fluids (e.g., water-carboxymethyl cellulose solutions). An example of liquid/liquid systems is also shown, in which case a non-Newtonian liquid of a more complex rheology (emulsion, with Newtonian infinity-viscosity at high shear rates) is considered. This has shown to have a prominent effect on the two-phase flow characteristics. The issue of the pressure gradient reduction is addressed focusing on the lubrication of the flow of a high viscous shear-thinning emulsion by adding water.

The implications of the exact solution on the modelling of the well-known and widely used one-dimensional Two-Fluid model are discussed. The solutions obtained for the velocity profiles were used to examine the effect of the liquid rheology on the velocity profile shape factors, which are needed to represent correctly the liquid inertia terms. It was shown that when the shear thinning effect is predominant, the shape factor values are lower compared to the corresponding values when a Newtonian behaviour is assumed (due to the tendency to form a plug-like shape of the velocity profile). An algorithm, which based on the Two-Fluid model predictions was presented and tested, to a priori predict when an asymptotic Newtonian (zero-viscosity, or infinity-viscosity) behaviour should be considered for specified flow conditions. Such a tool can be useful for avoiding large errors in predictions due to an incorrect viscosity model.

Since the majority of the experimental data available in the literature concern gas/liquid flows, the capability of the Two-Fluid model to predict the flow characteristics was tested. The effect of turbulence in the gas layer was also considered, and it was shown that the gas turbulence extends the range of operational conditions where the power-law model is a valid rheological model. Nevertheless, for carrying a rigorous stability analysis, for identifying the operational zone where smooth-stratified flow can be stable, the starting point for the analysis should be the solution of laminar flow both in the Newtonian and non-Newtonian phases combined with a valid rheological model for the non-Newtonian phase.

**Acknowledgements**

We would want to thank Prof. Paola Gervasio from the Università degli Studi di Brescia for the useful support with the Chebyshev collocation points approach for solving non-linear equations.

## Appendix A: Another algorithm for the laminar solutions with power-law fluids

In this Appendix, a variation of the algorithm given in Section 2.2 is presented for the case of Newtonian/power-law fluid. If the non-Newtonian fluid is a power-law fluid the dimensionless constitutive relations becomes

$$\tilde{\tau}_{H,zy} = \tilde{\varkappa}_H \left|\frac{d\tilde{u}_H}{d\tilde{y}}\right|^{n_H-1}\left(\frac{d\tilde{u}_H}{d\tilde{y}}\right), \quad \tilde{\tau}_{L,zy} = \tilde{\varkappa}_L \left|\frac{d\tilde{u}_L}{d\tilde{y}}\right|^{n_L-1}\left(\frac{d\tilde{u}_L}{d\tilde{y}}\right), \quad \tilde{\mu}^L = \frac{\varkappa_H\left|\frac{U_{Ls}}{H}\right|^{n_H-1}}{\mu_L}, \tag{A.1}$$

where $\tilde{\varkappa}_L = n_L = 1$, $n_H$ is the fluid behavior index of the heavy phase, and $\tilde{\varkappa}_H = \tilde{\mu}^L$; $\tilde{\mu}^L$ is the viscosity ratio of the system and the upper-script $L$ indicates that the viscosity ratio depends on the superficial velocity of the light phase. Obviously, the same algorithm presented in Section 2.2 can be used for this case using the constitutive relations, Eqs. (A.1), in Eq. (4). However, we present an alternative and more performing algorithm valid only for the case of a power-law fluid. The algorithm is based on the one proposed by Taghavi *et al.* (2009) and Alba *et al.* (2013).

The same structure of the algorithm given in Section 2.2 is maintained and the differences are the following: the inner loop (solving Eq. (4) with Broyden's algorithm) is not needed; as the interfacial velocities required in $\tilde{u}_{i,H,L}$ in the no-slip condition and the integrals of Eq. (5) are computed analytically. The idea is to write the shear stresses as a function of the position and the other dimensionless parameters as:

$$\tilde{\tau}_{zy,H}(\tilde{y}) = \tilde{\tau}_H\left(1-\frac{\tilde{y}}{\tilde{h}}\right) + \frac{\tilde{\tau}_i\tilde{y}}{\tilde{h}}, \quad \tilde{\tau}_{zy,L}(\tilde{y}) = \tilde{\tau}_L\left(\frac{\tilde{h}-\tilde{y}}{\tilde{h}-1}\right) + \frac{\tilde{\tau}_i(1-\tilde{y})}{1-\tilde{h}}, \tag{A.2}$$

where the shear stress of the heavy and light phase at the wall is $\tilde{\tau}_H(\tilde{\tau}_i, \tilde{P}_L, \tilde{h}) = \tilde{\tau}_i - 12\tilde{h}(\tilde{P}_L - Y_L)$ and $\tilde{\tau}_L(\tilde{\tau}_i, \tilde{P}_L, \tilde{h}) = \tilde{\tau}_i + (12\tilde{P}_L)(1-\tilde{h}))$. The velocity profile are computed by integrating the velocity gradient:

$$\frac{\partial \tilde{u}_{L,H}(\tilde{y})}{\partial \tilde{y}} = \mathrm{sgn}(\tilde{\tau}_{zy,L,H})\frac{|\tilde{\tau}_{zy,L,H}(\tilde{y})|^{1/n_{H,L}}}{\tilde{\varkappa}_{H,L}{}^{1/n_{H,L}}}, \tag{A.3}$$

yielding

$$\tilde{u}_H(\tilde{y}) = \int_0^{\tilde{y}} \frac{\partial \tilde{u}_L(\tilde{y})}{\partial \tilde{y}} d\tilde{y} = \frac{\tilde{h}(|\tilde{\tau}_{zy,H}(\tilde{y})|^{1/n_H+1} - (|\tilde{\tau}_H|^{1/n_H+1})}{(1/n_H+1)\tilde{\varkappa}_H^{1/n_H}(\tilde{\tau}_i-\tilde{\tau}_H)}, \tag{A.4a}$$

$$\tilde{u}_L(\tilde{y}) = \int_{\tilde{y}}^{1} \frac{\partial \tilde{u}_L(\tilde{y})}{\partial \tilde{y}} d\tilde{y} = -\frac{(1-\tilde{h})(|\tilde{\tau}_{zy,L}(\tilde{y})|^{1/n_L+1} - (|\tilde{\tau}_L|^{1/n_L+1})}{(1/n_L+1)\tilde{\varkappa}_L^{1/n_L}(\tilde{\tau}_i-\tilde{\tau}_L)}, \tag{A.4b}$$

Therefore, the interfacial velocities are

$$\tilde{u}_{iH} = \int_0^{\tilde{h}} \frac{\partial \tilde{u}_H(\tilde{y})}{\partial \tilde{y}} d\tilde{y} = \begin{cases} \frac{\tilde{h}\left(|\tilde{\tau}_i|^{1/n_H+1} - |\tilde{\tau}_H|^{1/n_H+1}\right)}{\tilde{\varkappa}_H^{1/n_H+1}(1/n_H+1)(\tilde{\tau}_i-\tilde{\tau}_H)} & \tilde{\tau}_i \neq \tilde{\tau}_H, \\ \frac{\tilde{h}\,\mathrm{sgn}(\tilde{\tau}_i)\left(|\tilde{\tau}_i|^{1/n_H}\right)}{\tilde{\varkappa}_H^{1/n_H}} & \tilde{\tau}_i = \tilde{\tau}_H, \end{cases} \tag{A.5a}$$

$$\tilde{u}_{iL} = \int_{\tilde{h}}^{1} \frac{\partial \tilde{u}_L(\tilde{y})}{\partial \tilde{y}} d\tilde{y} = \begin{cases} \frac{(\tilde{h}-1)\left(|\tilde{\tau}_i|^{1/n_L+1} - |\tilde{\tau}_L|^{1/n_L+1}\right)}{\tilde{\varkappa}_L^{1/n_L+1}(1/n_L+1)(\tilde{\tau}_i-\tilde{\tau}_L)} & \tilde{\tau}_i \neq \tilde{\tau}_L, \\ \frac{(\tilde{h}-1)\,\mathrm{sgn}(\tilde{\tau}_i)\left(|\tilde{\tau}_i|^{1/n_L}\right)}{\tilde{\varkappa}_L^{1/n_L}} & \tilde{\tau}_i = \tilde{\tau}_L, \end{cases} \tag{A.5b}$$

The analytical expressions for the integrals in the flow rate constrains (Eq. 5) are:

$$\int_0^{\tilde{h}} \tilde{u}_H(\tilde{y})\, d\tilde{y} = \begin{cases} \tilde{h}\,\tilde{u}_{iH} - \frac{\tilde{h}^2}{\tilde{\varkappa}_H^{1/n_H}(\tilde{\tau}_i-\tilde{\tau}_H)^2}\left[\frac{\mathrm{sgn}(\tilde{\tau}_H)|\tilde{\tau}_H|^{1/n_H+2}}{(1/n_H+1)(1/n_H+2)} + \right. \\ \left. +\mathrm{sgn}(\tilde{\tau}_i)|\tilde{\tau}_i|^{1/n_H+1}\left(\frac{\tilde{\tau}_i-\tilde{\tau}_H}{(1/n_H+1)\mathrm{sgn}(\tilde{\tau}_i)} - \frac{|\tilde{\tau}_i|}{(1/n_H+1)(1/n_H+2)}\right)\right] & \tilde{\tau}_i \neq \tilde{\tau}_H, \\ \tilde{h}\,\tilde{u}_{iH} - \frac{\tilde{h}^2\mathrm{sgn}(\tilde{\tau}_i)|\tilde{\tau}_i|^{1/n_H}}{2\tilde{\varkappa}_H^{1/n_H}} & \tilde{\tau}_i = \tilde{\tau}_H, \end{cases} \tag{A.6a}$$

$$\int_{\tilde{h}}^{1} \tilde{u}_L(\tilde{y})\, d\tilde{y} = \begin{cases} (1-\tilde{h})\,\tilde{u}_{iL} - \frac{(1-\tilde{h})^2}{\tilde{\varkappa}_L^{1/n_L}(\tilde{\tau}_i-\tilde{\tau}_L)^2}\left[\frac{\mathrm{sgn}(\tilde{\tau}_L)|\tilde{\tau}_L|^{1/n_L+2}}{(1/n_L+1)(1/n_L+2)} + \right. \\ \left. +\mathrm{sgn}(\tilde{\tau}_i)|\tilde{\tau}_i|^{1/n_L+1}\left(\frac{\tilde{\tau}_i-\tilde{\tau}_L}{(1/n_L+1)\mathrm{sgn}(\tilde{\tau}_i)} - \frac{|\tilde{\tau}_i|}{(1/n_L+1)(1/n_L+2)}\right)\right] & \tilde{\tau}_i \neq \tilde{\tau}_L, \\ (1-\tilde{h})\tilde{u}_{iL} - \frac{(1-\tilde{h})^2\mathrm{sgn}(\tilde{\tau}_i)|\tilde{\tau}_i|^{1/n_L}}{2\tilde{\varkappa}_L^{1/n_L}} & \tilde{\tau}_i = \tilde{\tau}_L, \end{cases} \tag{A.6b}$$

We presented here the algorithm for the case of heavy non-Newtonian/light Newtonian configurations, which can be easily modified for the heavy-Newtonian/light non-Newtonian flow configuration.

## Appendix B. Friction factor for a power-law fluid in a two-plate geometry.

The friction factor for a single-phase power-law fluid is reviewed by Chhabra and Richardson (2008) for the pipe geometry and more recently for open channel with different cross sectional geometry by Burger *et al.* (2010). The formulation of the Reynolds number, which appears in the friction factor formulae, is the well-known Metzner and Reed's Reynolds number for power-law fluid in a pipe, see Metzner and Reed (1959).

The laminar single-phase (superficial) friction factor for a power-law fluid in a two-plate geometry (based on the hydraulic diameter) is not reported in the literature and it can be easily obtained in terms of the hydraulic diameter

$$f = \frac{24}{Re}, \quad Re = \frac{\rho_j U_{js}^{2-n} {D_h}^n}{\kappa_j 12^{n-1}.\left(\frac{1+2n}{3n}\right)^n} = \frac{\rho_j U_{js} D_h}{\kappa_j\left(\frac{12\,U_{js}}{D_h}\right)^{n-1}\left(\frac{1+2n}{3n}\right)^n} = \frac{\rho_j U_{js} D_h}{\mu_{js,eff}}, \quad D_h = 2H, \quad \mu_{js,eff} = \kappa_j\left(\frac{12\,U_{js}}{D_h}\right)^{n-1}\left(\frac{1+2n}{3n}\right)^n, \tag{A.7}$$

where $H$ is the channel depth, $\rho_j$ is the density, $U_{js}$ is the (superficial) velocity, $n$ is the fluid behavior index, and $\kappa_j$ is the fluid consistency index of the $j$-phase, $\mu_{js,eff}$ is the effective viscosity. The Reynolds number for a two-plate geometry is similar to the well-known Metzner and Reed Reynolds number for the pipe geometry, the only differences are due to the different shear rate at the wall in a two-plate geometry $\dot{\gamma}_{ws} = \frac{12U}{D_h}\frac{1+2n}{3n}$. Note that the Reynolds number converge to Newtonian one if $n=1$.

## Appendix C. TF SP-PL model a Newtonian/power-law fluid in a two-plate geometry.

The well-known Two-Fluid model has been proposed for shear-thinning power-law fluid by Heywood and Charles (1979) and Picchi *et al.* (2014). In two-plate geometry the combined momentum equation for the steady state and fully developed flow has to be satisfied for a give flow conditions, yielding

$$\frac{\tau_H S_H}{A_H} - \frac{\tau_L S_L}{A_L} - \tau_i S_i\left(\frac{1}{A_H} + \frac{1}{A_L}\right) + (\rho_H - \rho_L) g \sin(\beta) = 0, \tag{A.8}$$

where $S_H = S_L = S_i = 1, A_H = h, A_H = H - h.$ The shear stress closures are based on the single-phase closure relations for a Power-law fluid (TF SP-PL):

$$\tau_H = -\frac{1}{2}\rho_H f_H U_H |U_H|, \tag{A.9a}$$

$$\tau_L = -\frac{1}{2}\rho_L f_L U_L |U_L|, \tag{A.9b}$$

$$\tau_i = -\frac{1}{2}\rho_i f_i (U_L - U_H)|U_L - U_H|\, F_{iw}, \tag{A.9c}$$

where the friction factor of the heavy non-Newtonian phase $f_H$ is given by Eq. (A.7) based on $U_H$ and the hydraulic diameter $D_H$; $f_L$ is given by Eq. (A.7) based on $U_L$ and the hydraulic diameter $D_L$ and assuming $n = 1$. The interfacial friction factor $f_i$ is $f_i = f_L$ if $U_L > U_H$ or $f_i = f_H$ otherwise; $F_{iw}$ is the augmentation factor due to waviness and it is assumed equal to unity in this work. The hydraulic diameters are calculated for gas/liquid systems, assuming the interface as a free surface for the liquid and as a wall for the gas, giving $D_H = 4h$ and $D_L = 2(H - h)$, while for liquid/liquid systems $D_H = 4h$ and $D_L = 2(H - h)$ if $U_L > U_H$ and $D_H = 2h$ and $D_L = 4(H - h)$ if $U_H > U_L$. The pressure gradient $dp/dz$ can be obtained from one of the two momentum equations yielding

$$dp/dz = \frac{\tau_L S_L}{A_L} + \frac{\tau_i S_i}{A_L} + \rho_L g \sin(\beta), \tag{A.10}$$

## Appendix D. Solution for turbulent gas/laminar shear-thinning liquid stratified flows.

When the gas phase is turbulent, the model and the computation algorithm have to account for some modifications with respect to the algorithm presented in Section 2.2. The momentum balance in the gas, Eq. (4b), is not solved, while the gas phase velocity (the average velocity $U_L$) and the pressure gradient ($dp/dz$, Eq. A.10) are computed accordingly to the two-fluid model in the gas layer, see Appendix C. This choice is reasonable and not introduces significant errors for the case of a turbulent phase (considering that also the MTF (Ullmann *et al.* (2004)) closure relation correction factors reduce to unity when the light phase is much faster than the liquid). From the computational point of view, only the flow constrains of the heavy phase (Eq. (5a)) has to be satisfied, while the inner loop reduces to the search of the liquid velocity profile for a given interfacial shear-stress $\tau_i$ (Eq. (A.9c)). Note that the normalization should be based on the single phase of the turbulent Newtonian gas, $(-dp_f\ /\ dz\ )_{Ls}$, and the viscosity ratio of the system in Eq. (4a) should be modified as

$$\tilde{\mu} = \frac{\mu_0}{12\rho_L H U_{Ls} f_{Ls}}, \quad (-dp_f\ /\ dz\ )_{Ls} = \frac{\rho_L U_{Ls}^2 f_{Ls}}{H}. \tag{A.11}$$

The turbulent friction factor $f_{turb}$ is given by (White (2008))

$$\frac{1}{\sqrt{f_{turb}/4}} = 2\log_{10}\left(Re_{D_h}\sqrt{f_{turb}\ /\ 4}\right) - 1.19, \tag{A.12}$$

where $Re_{D_h}$ is the Reynolds number based on the hydraulic diameter of the layer and its insitu average phase velocity, which is equal to the light phase superficial Reynolds number is $Re_{Ls} = \rho_L U_{Ls} D_h/\mu_L.$, $D_h = 2H$.

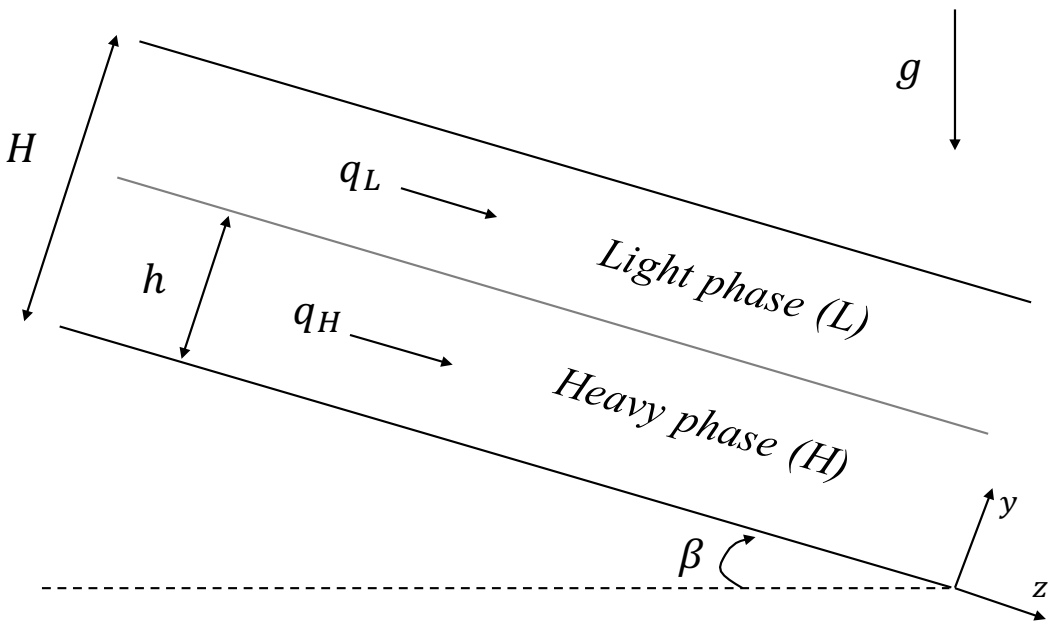

**Figure 1**. Schematic description of the stratified flow configuration and coordinates.

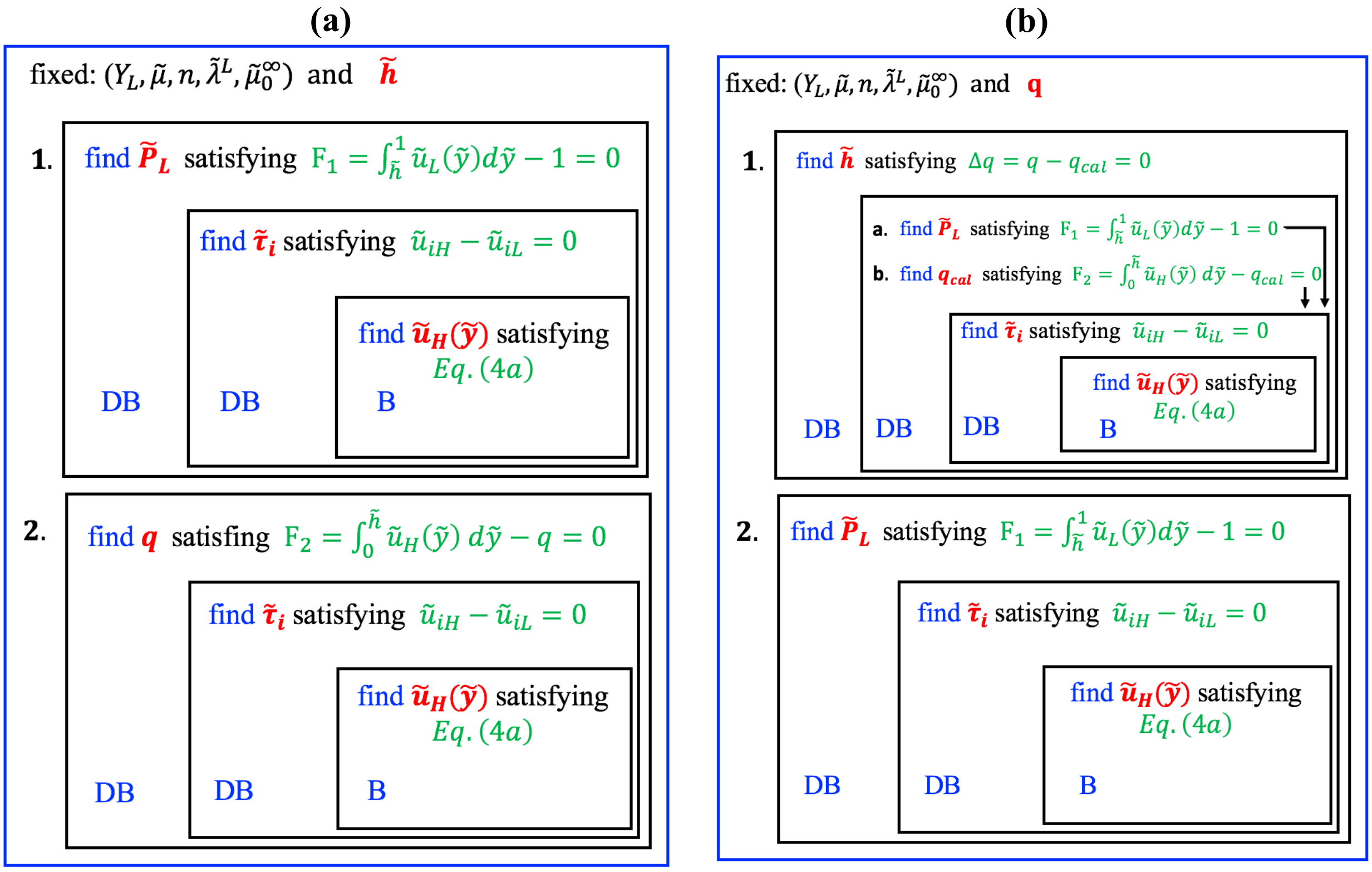

**Figure 2**. Block diagram of the algorithm to compute the laminar solutions: (a) for a given holdup and (b) for a given flow rate ratio. Each black box represents an iteration loop to find the variable (in red). DB stays for Dekker and Brent's algorithm, B stays for Broyden's algorithm.

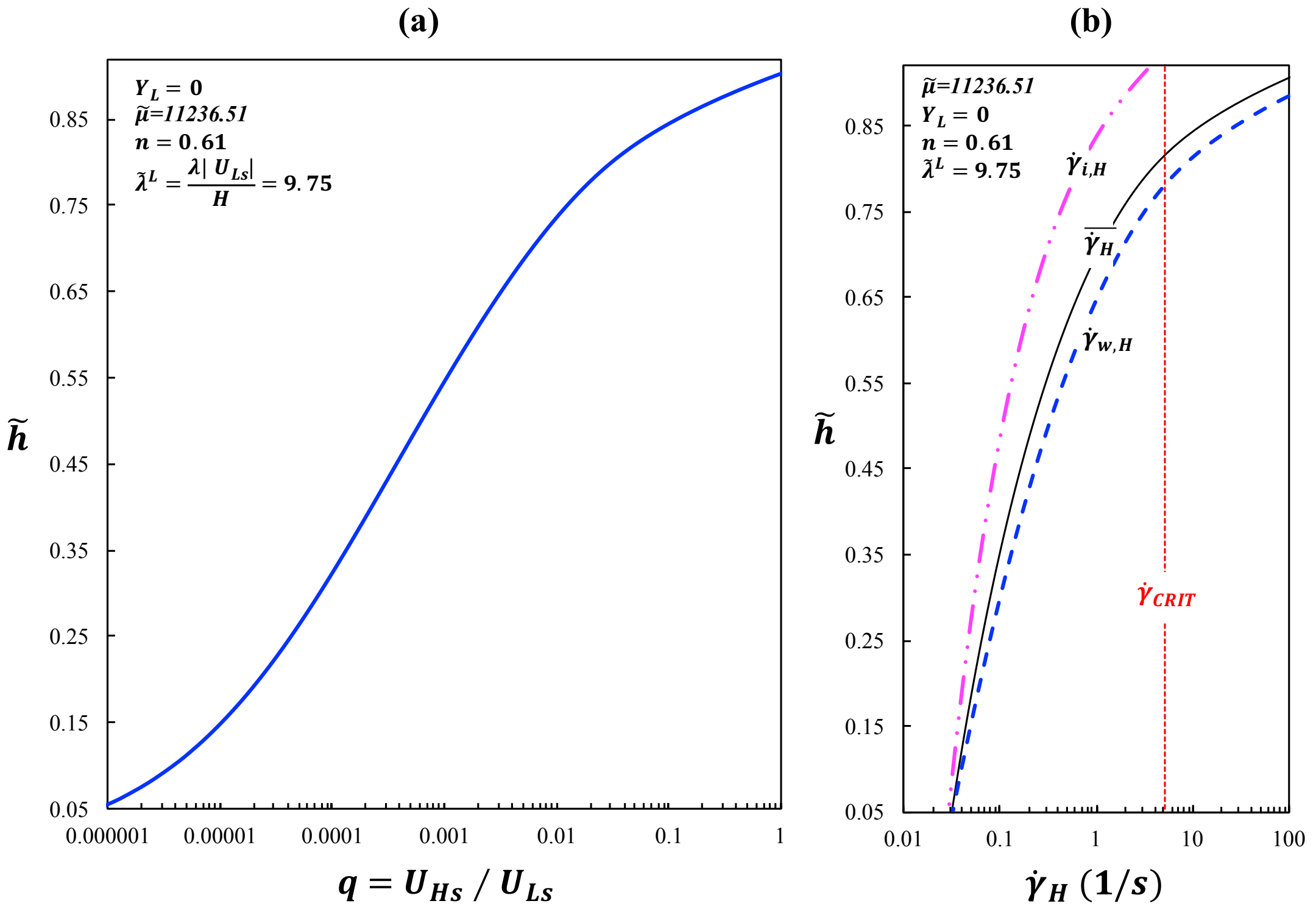

**Figure 3.** (a) Exact holdup curve for horizontal gas/shear-thinning fluid stratified flow (Carreau fluid); (b) Shear rate at the wall and at the interface, average shear rate of the shear-thinning fluid as a function of the holdup. The combinations of nondimensional parameters corresponds to a 2 cm height channel, $U_{LS} = 1$ m/s, and the shear-thinning fluid is CMC05.

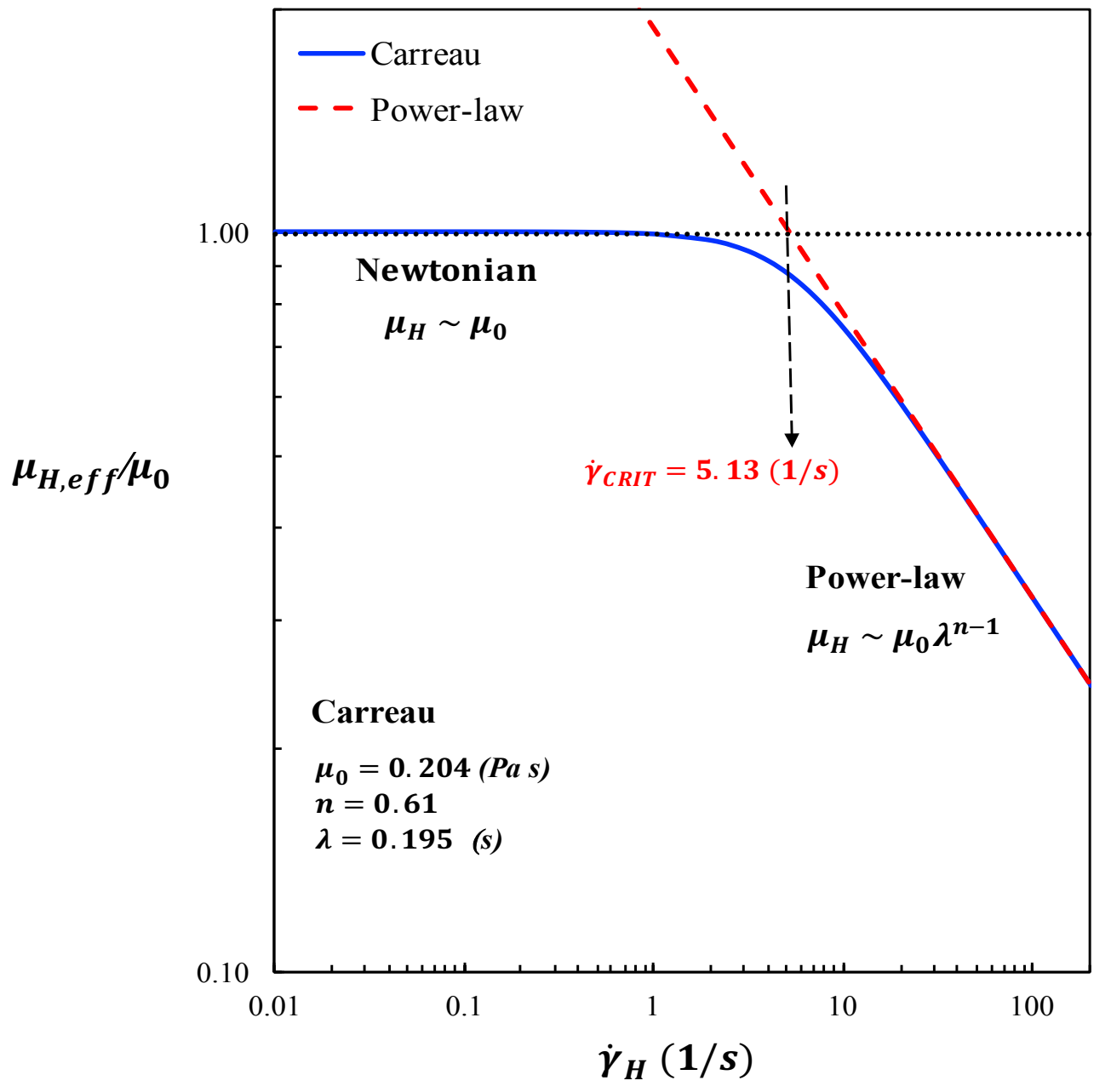

**Figure 4**. Viscosity curves for the CMC05 shear-thinning fluid: the Carreau viscosity model and the two asymptotic viscosity models (the zero-viscosity Newtonian and the power-law) are plotted. $\dot{\gamma}_{CRIT}$ is the shear rate which corresponds to the intersection of the two asymptotic behaviours.

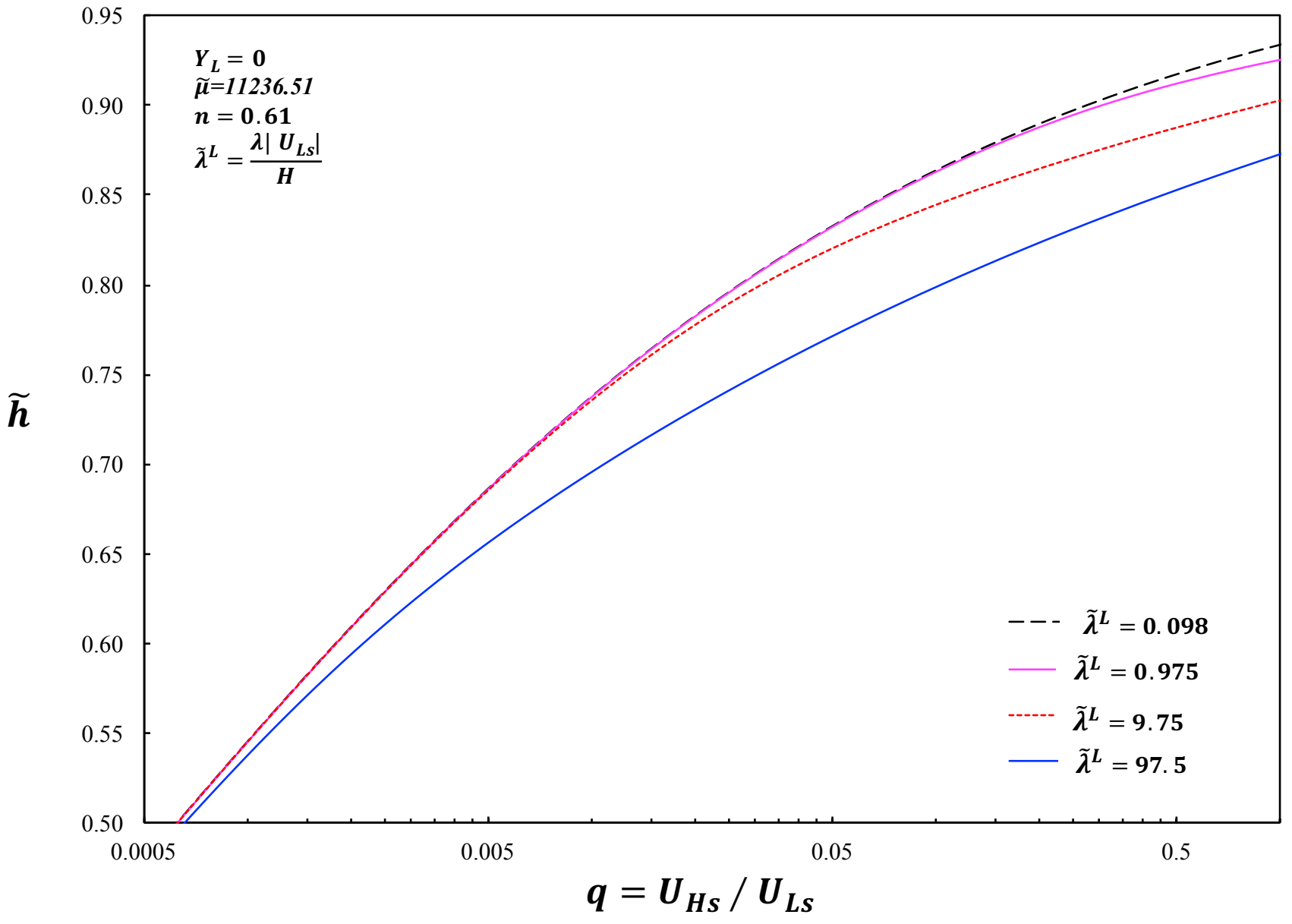

**Figure 5.** Effect of the dimensionless time constant on the holdup curve for horizontal gas/shear-thinning fluid flows. The combinations of nondimensional parameters corresponds to a 2 cm height channel, and the shear-thinning fluid is CMC05

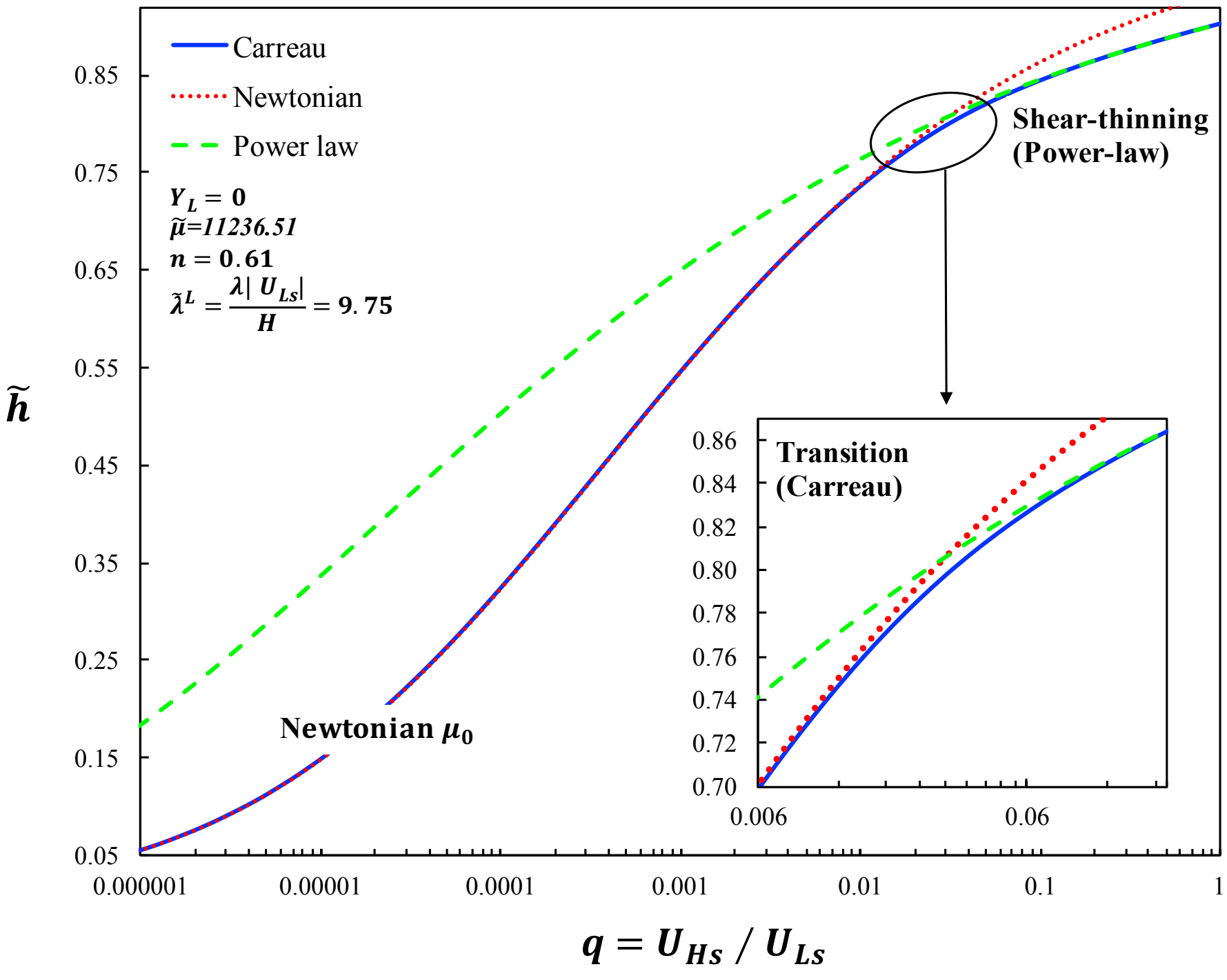

**Figure 6.** Asymptotic hold-up curves (Newtonian zero-viscosity and power-law behaviour) plotted with the exact holdup curve for horizontal gas/shear-thinning fluid stratified flow. The combinations of nondimensional parameters corresponds to a 2 cm height channel, $U_{LS} = 1$ m/s, and the shear-thinning fluid is CMC05.

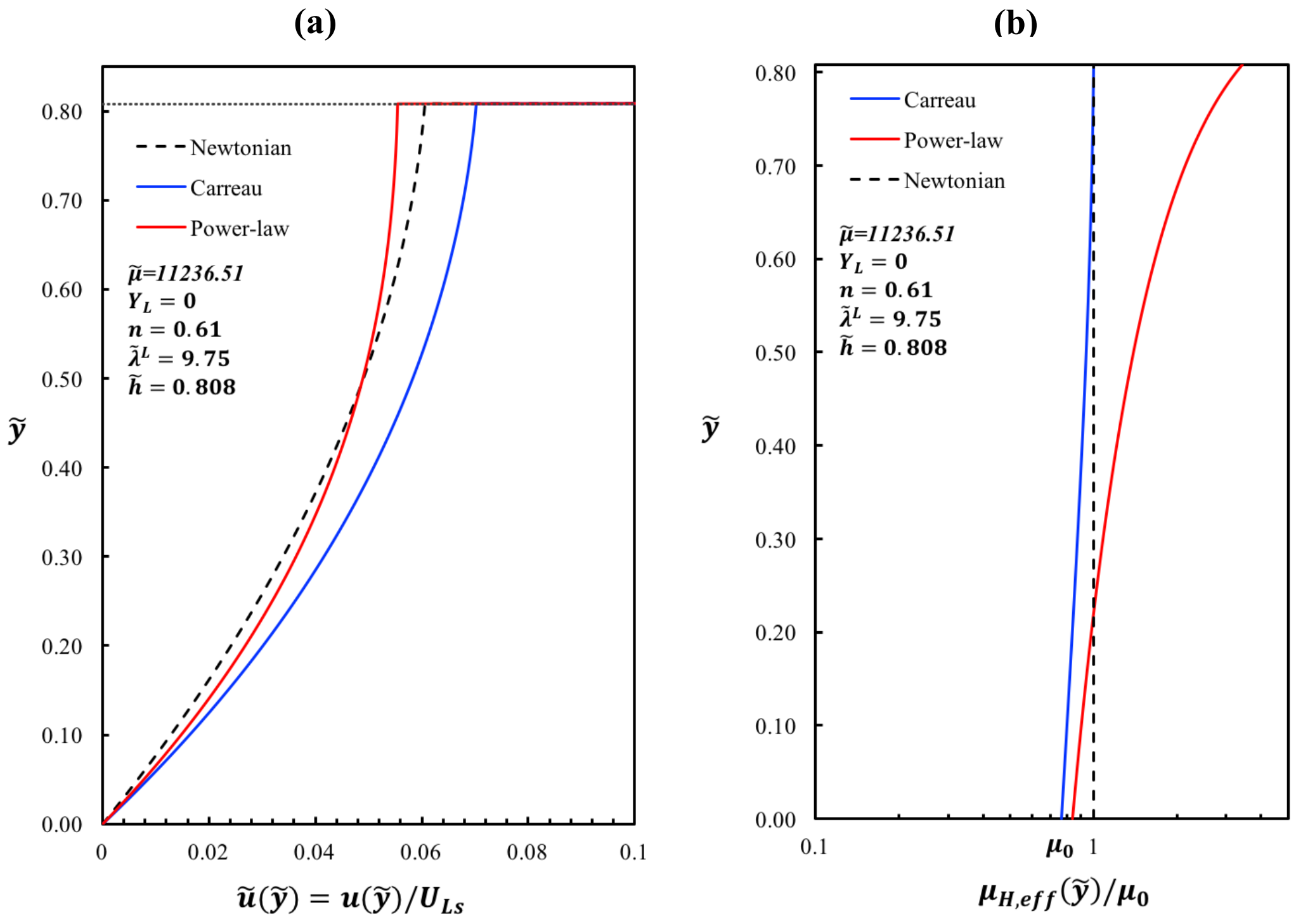

**Figure 7**. Exact and asymptotic velocity profiles (a) and effective viscosity distribution (b) in the liquid shear-thinning layer for horizontal gas/liquid stratified flow corresponding to the asymptotes intersection in Fig. 4. The combinations of nondimensional parameters corresponds to a 2 cm height channel, $U_{LS} = 1$ m/s, and the shear-thinning fluid is CMC05.

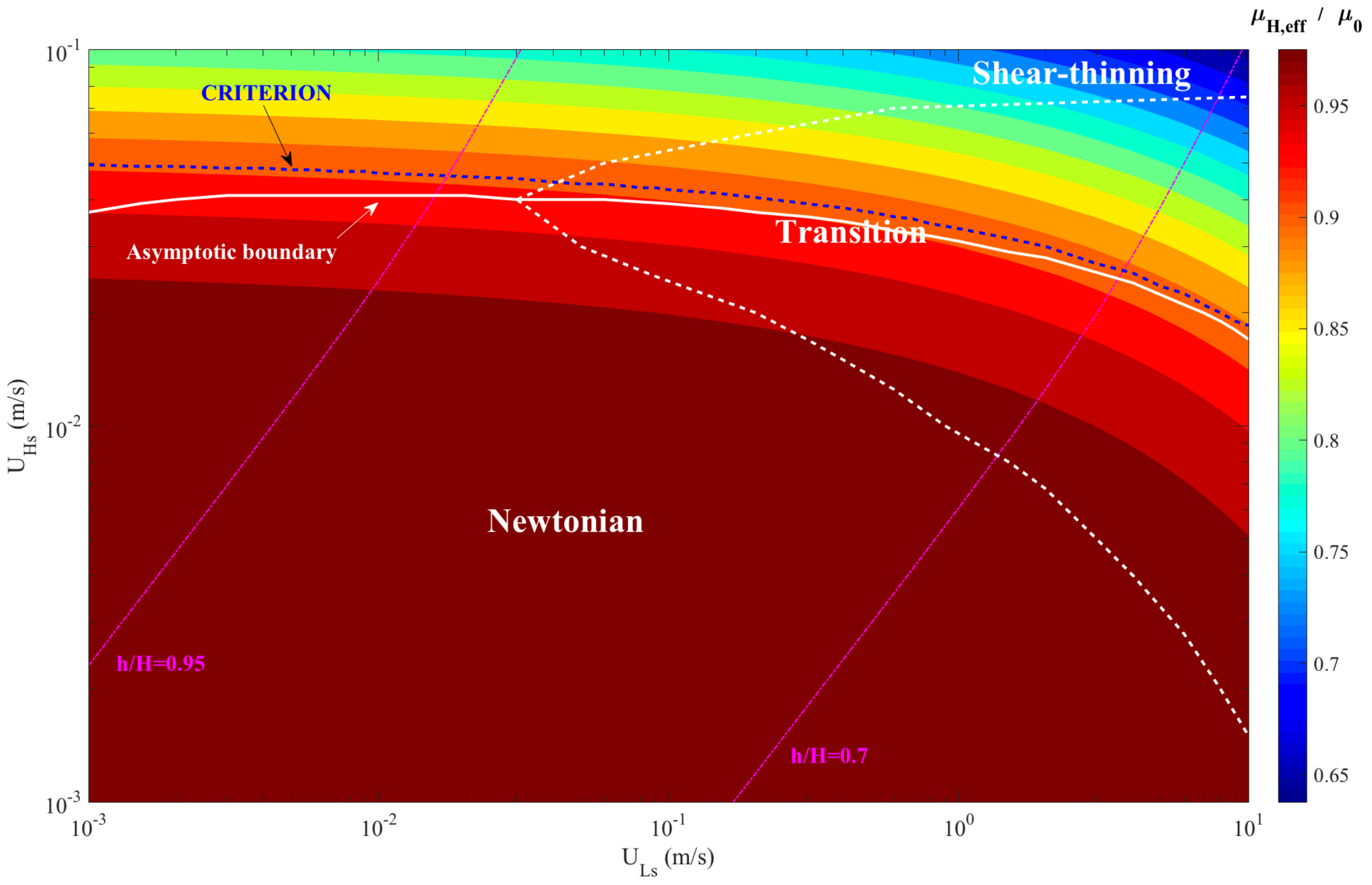

**Figure 8**. Contours of the average effective viscosity in the non-Newtonian layer plotted is superficial velocities coordinates for horizontal air/CMC05 system in a 2cm height channel. The rheological regions, the iso-holdup curves (pink), and the asymptotic boundary (white solid), which corresponds to the CRIT conditions of Fig. 4, are plotted. The CRITERION line refers to the switch between the Newtonian and the shear-thinning behaviour predicted by the two-fluid model, see Section 4.2.

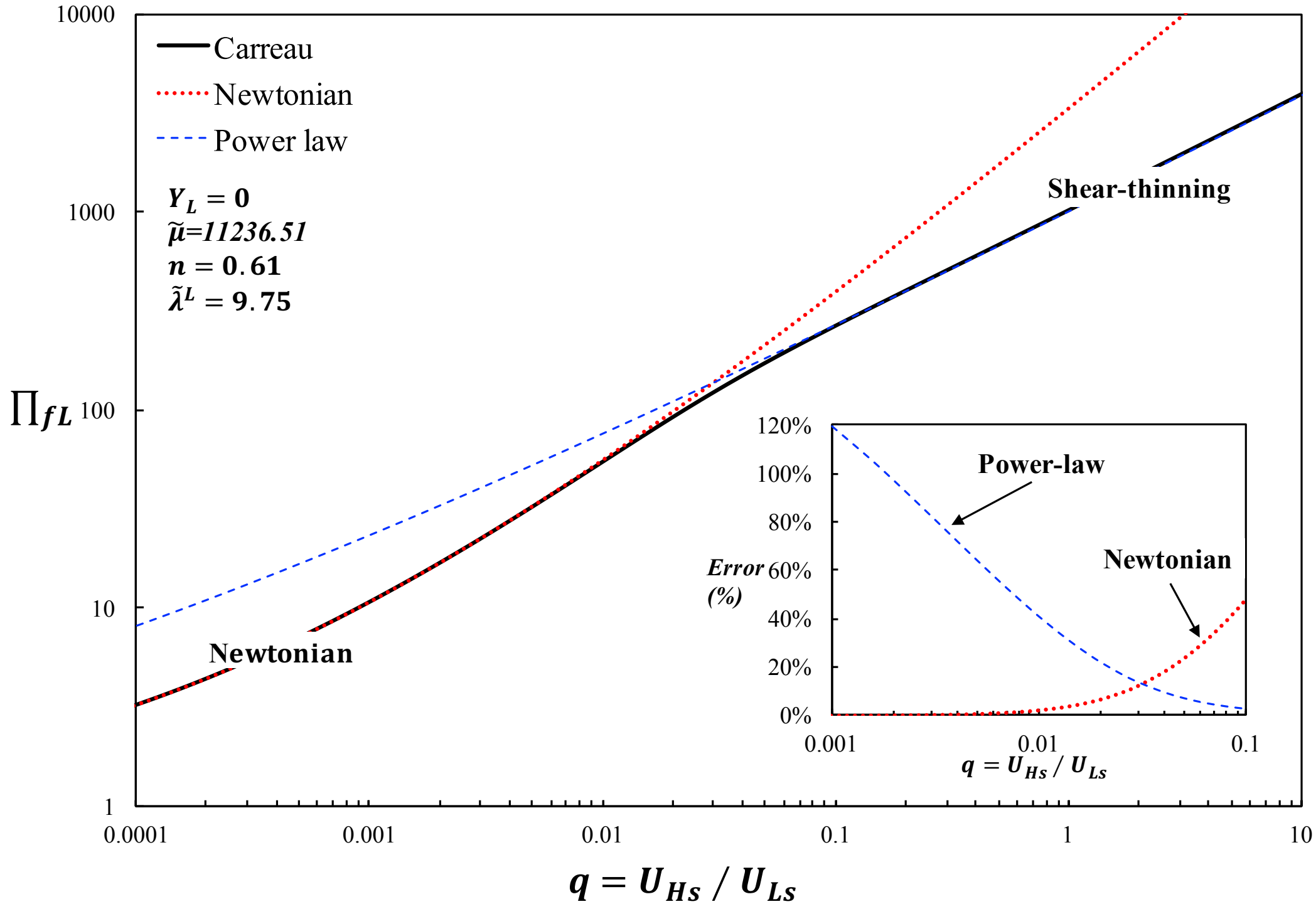

**Figure 9**. Frictional pressure gradient curve for horizontal gas/shear-thinning fluid stratified flow (Carreau fluid); the asymptotic hold-up curves (Newtonian zero-viscosity and power-law behaviour) are also plotted. The error introduced by using an incorrect viscosity model is shown. The combinations of nondimensional parameters corresponds to a 2 cm height channel, $U_{LS} = 1$ m/s, and the shear-thinning fluid is CMC05.

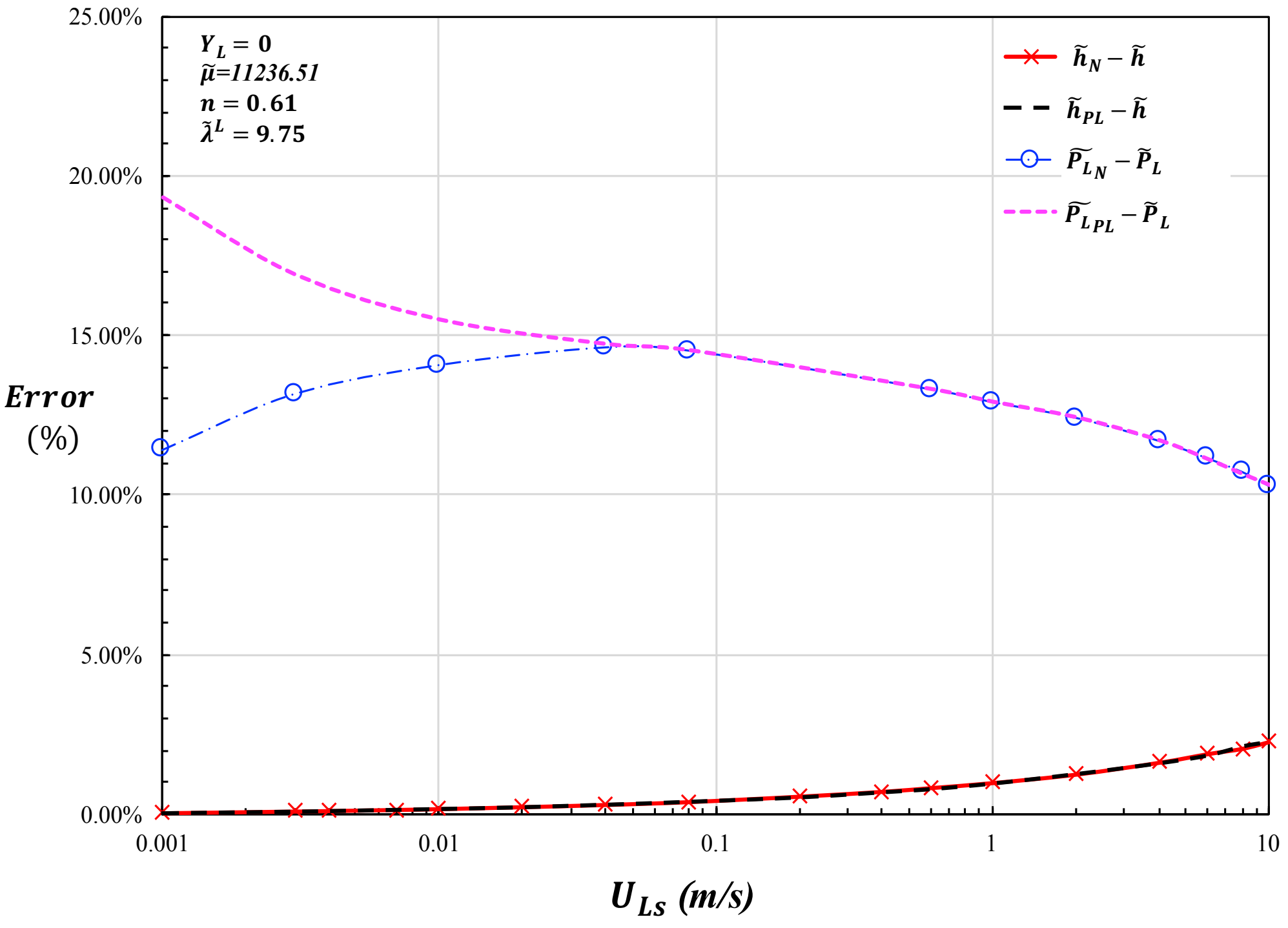

**Figure 10.** Error between exact solution (Carreau) and the asymptotes in hold-up and pressure gradient predictions along the asymptotic boundary of Fig 8. The combinations of nondimensional parameters corresponds to a 2 cm height channel, $U_{LS} = 1$ m/s, and the shear-thinning fluid is CMC05.

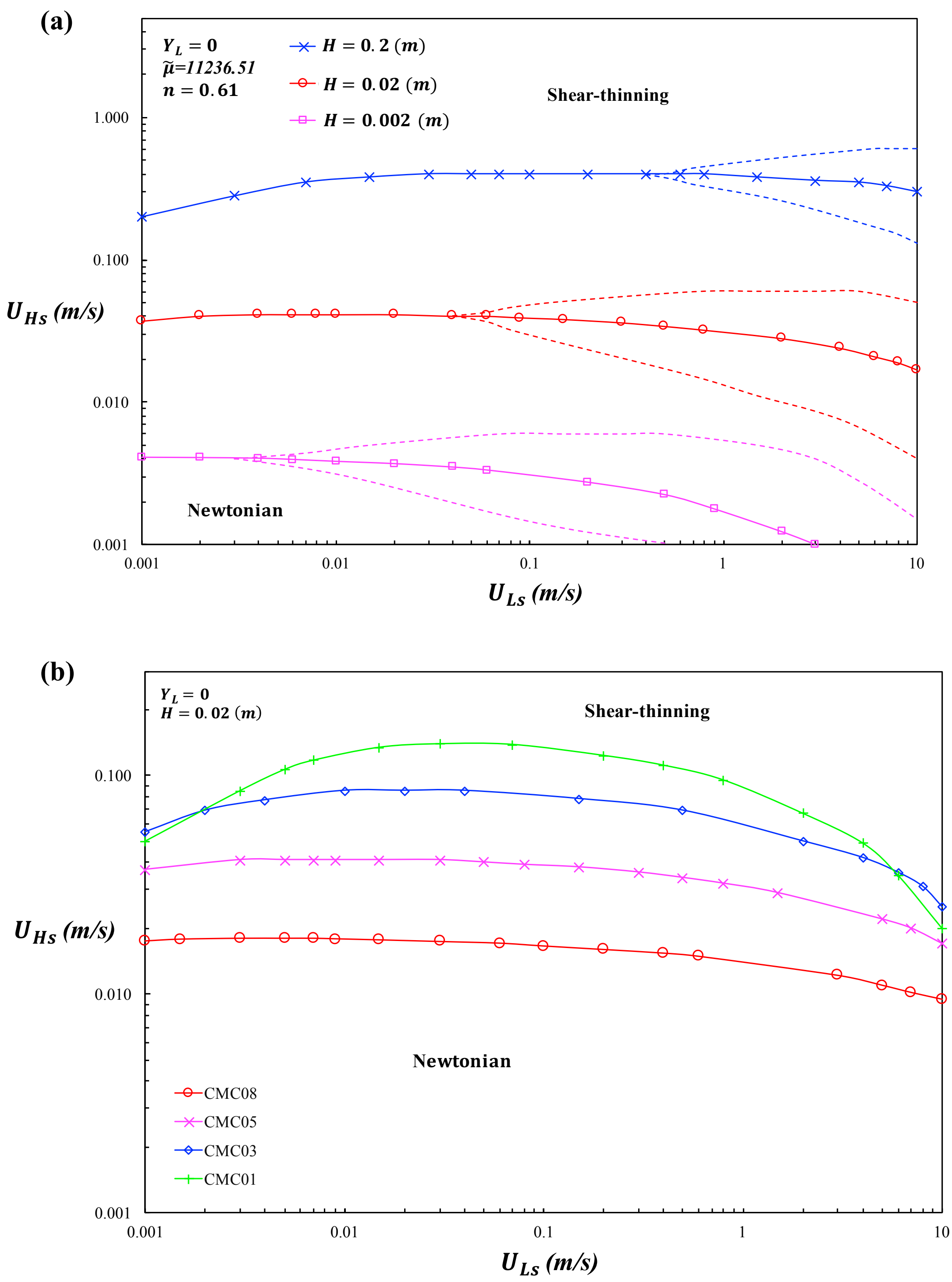

**Figure 11.** (a) Effect of channel size on the rheological regions for air/CMC05 systems: solid line represents the asymptotic boundary, dotted lines delimit the transition region. (b) Effect of the non-Newtonian liquid rheology on the rheological regions in the flow map for horizontal flow: the asymptotic boundary for different CMC solution is plotted for a fixed channel size.

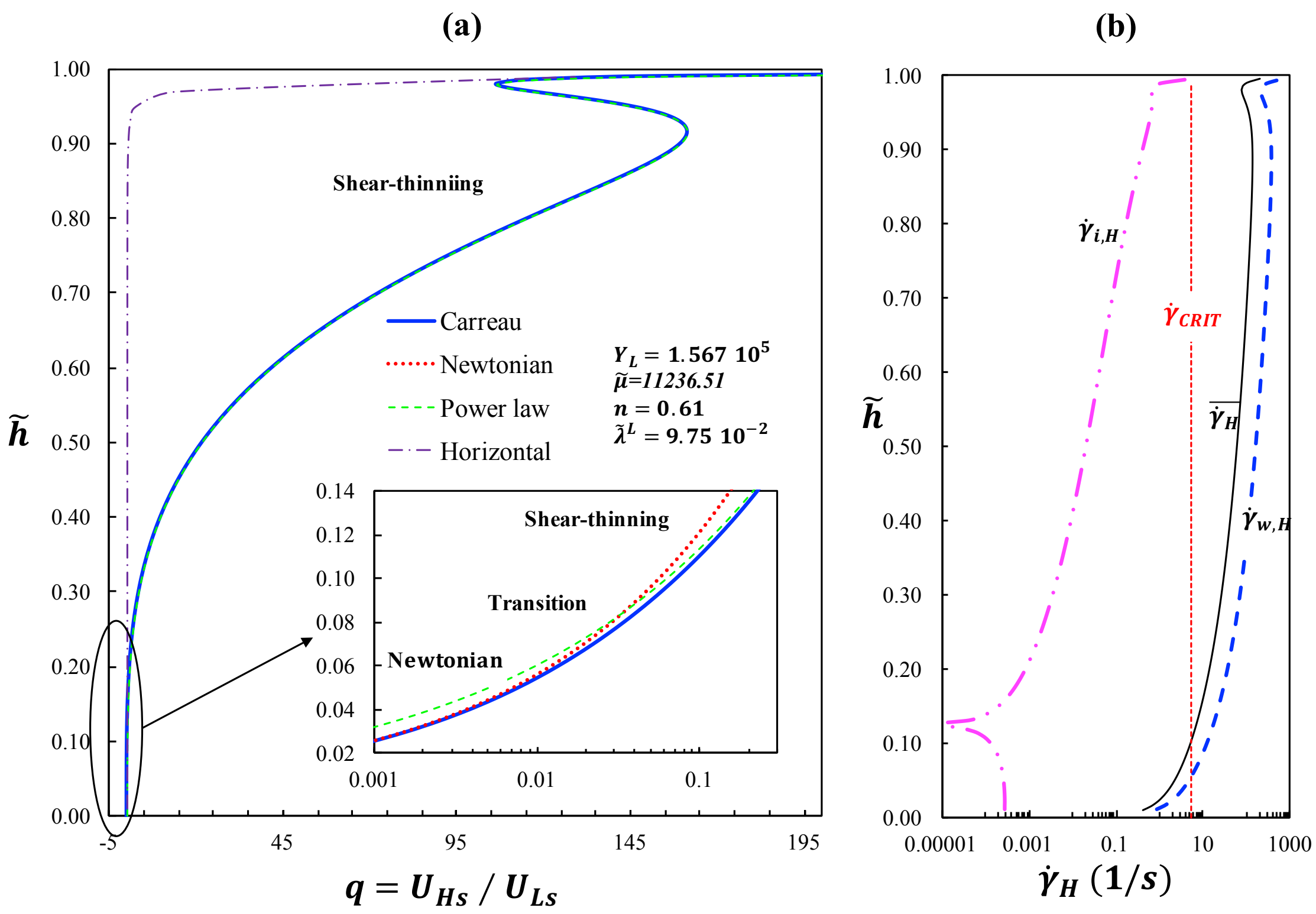

**Figure 12**. (a) Exact holdup curve for downward inclined gas/shear-thinning fluid stratified flow (Carreau fluid); the holdup curve corresponding to horizontal flow are also plotted. Asymptotic hold-up curves (Newtonian zero-viscosity and power-law behaviour) are also plotted. (b) Shear rate at the wall and at the interface, average shear rate of the shear-thinning fluid as a function of the holdup. The combinations of nondimensional parameters corresponds to a 2 cm height channel, $U_{LS} = 0.01$ m/s, the inclination angle is $\beta = 5°$ (downward), and the shear-thinning fluid is CMC05.

**(a)**

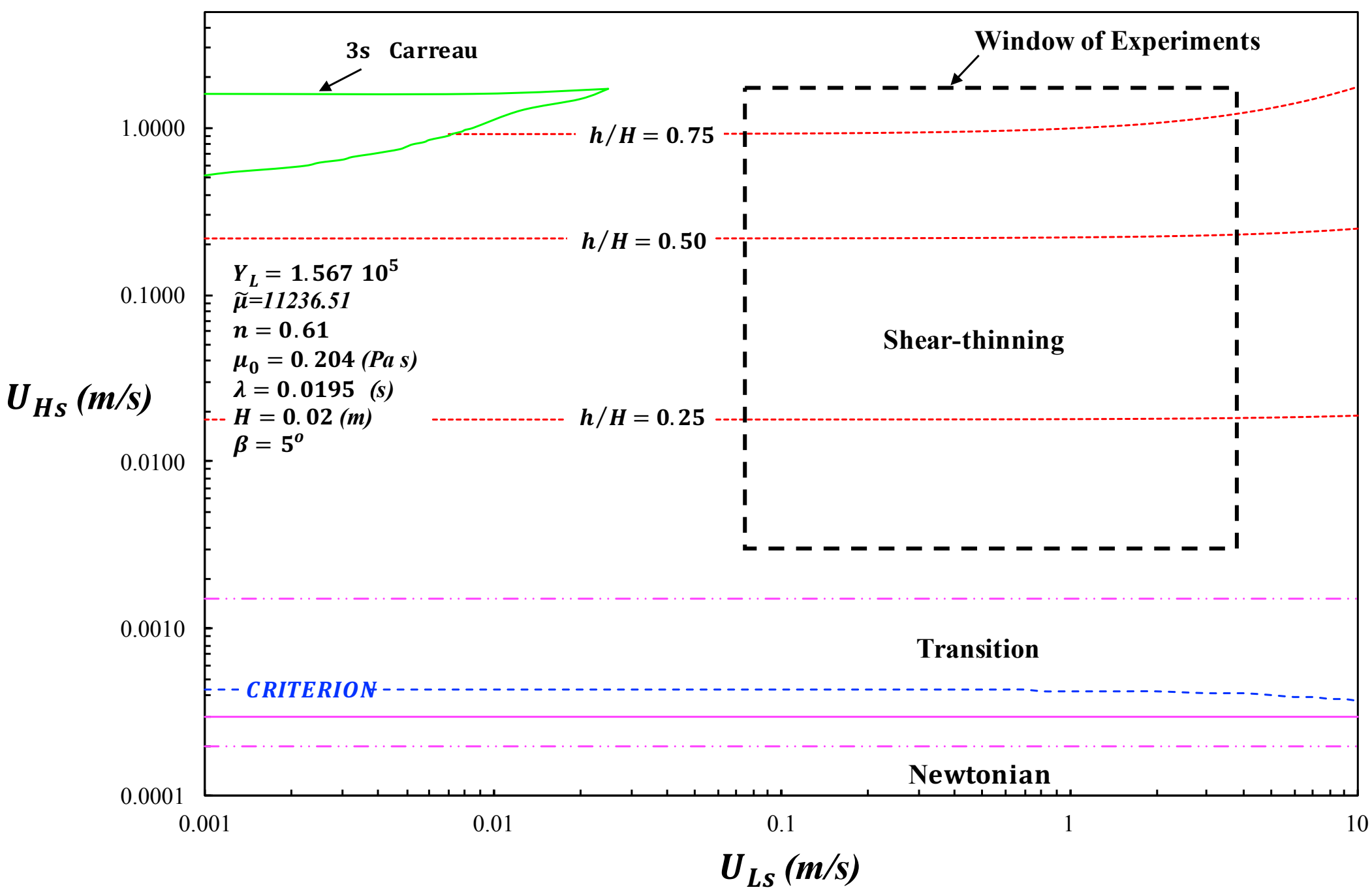

**(b)**

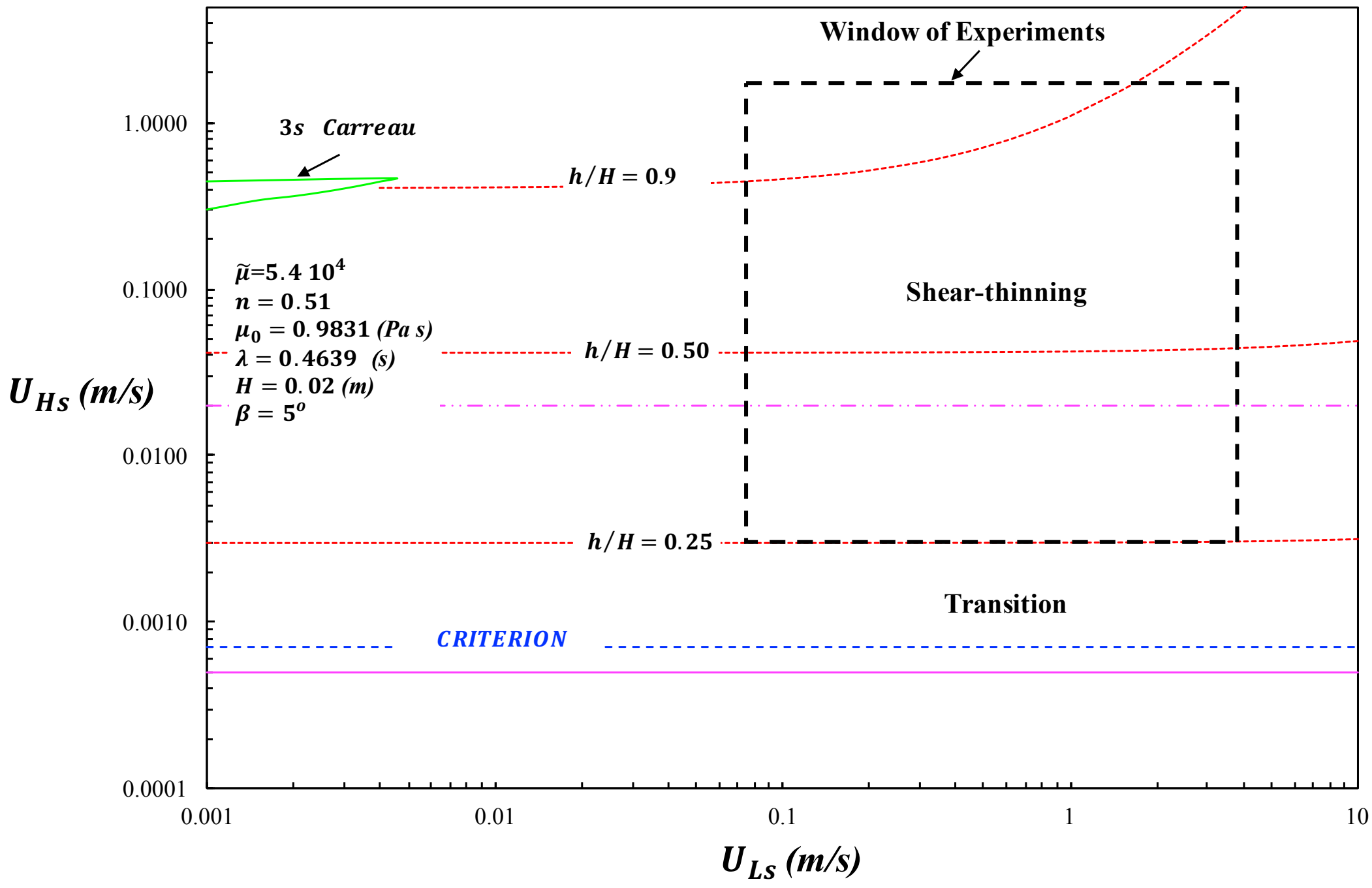

**Figure 13**. Flow pattern map for downward inclined gas/shear-thinning fluid flows: air/ CMC05 solution, (a), and air CMC08 solution, (b). The multiple solutions boundaries, the asymptotic boundary (solid lines), the rheological regions (delimited by dotted lines), and the iso-holdup curve are plotted. The windows of experiments corresponds to conditions where the stratified flow was observed experimentally (Tab. 2). The CRITERION line refers to the switch between the Newtonian and the power-law behaviour predicted by the two-fluid model, see Section 4.2.

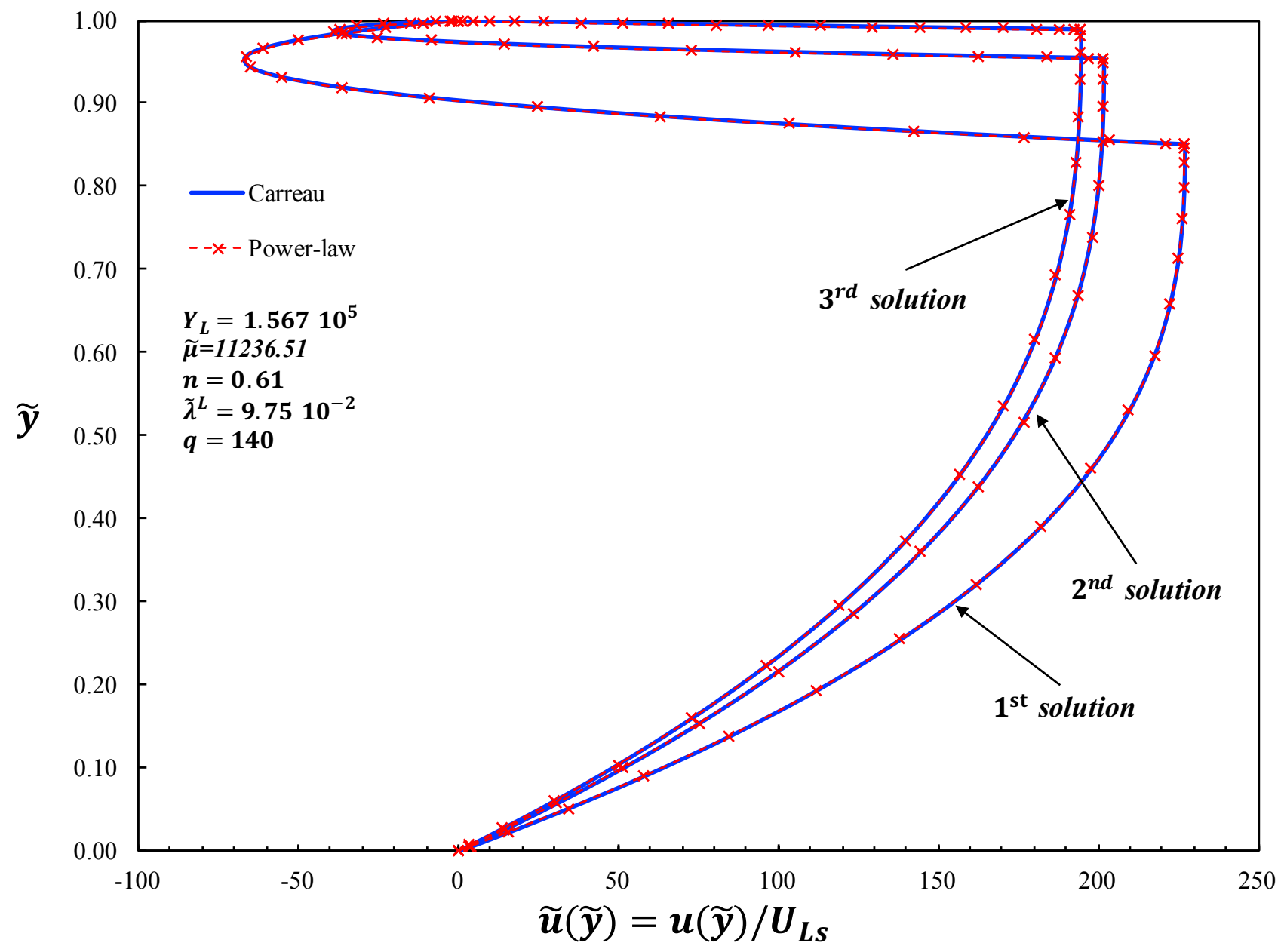

**Figure 14**. Velocity profile associated to triple solution in concurrent downward inclined flows for gas/shear-thinning fluid. The exact (Careau fluid) solution and the power law asymptotic solution are plotted.The combinations of nondimensional parameters corresponds to a 2 cm height channel, $U_{LS} = 0.01$ m/s, the inclination angle is $\beta = 5°$ (downward), and the shear-thinning fluid is CMC05.

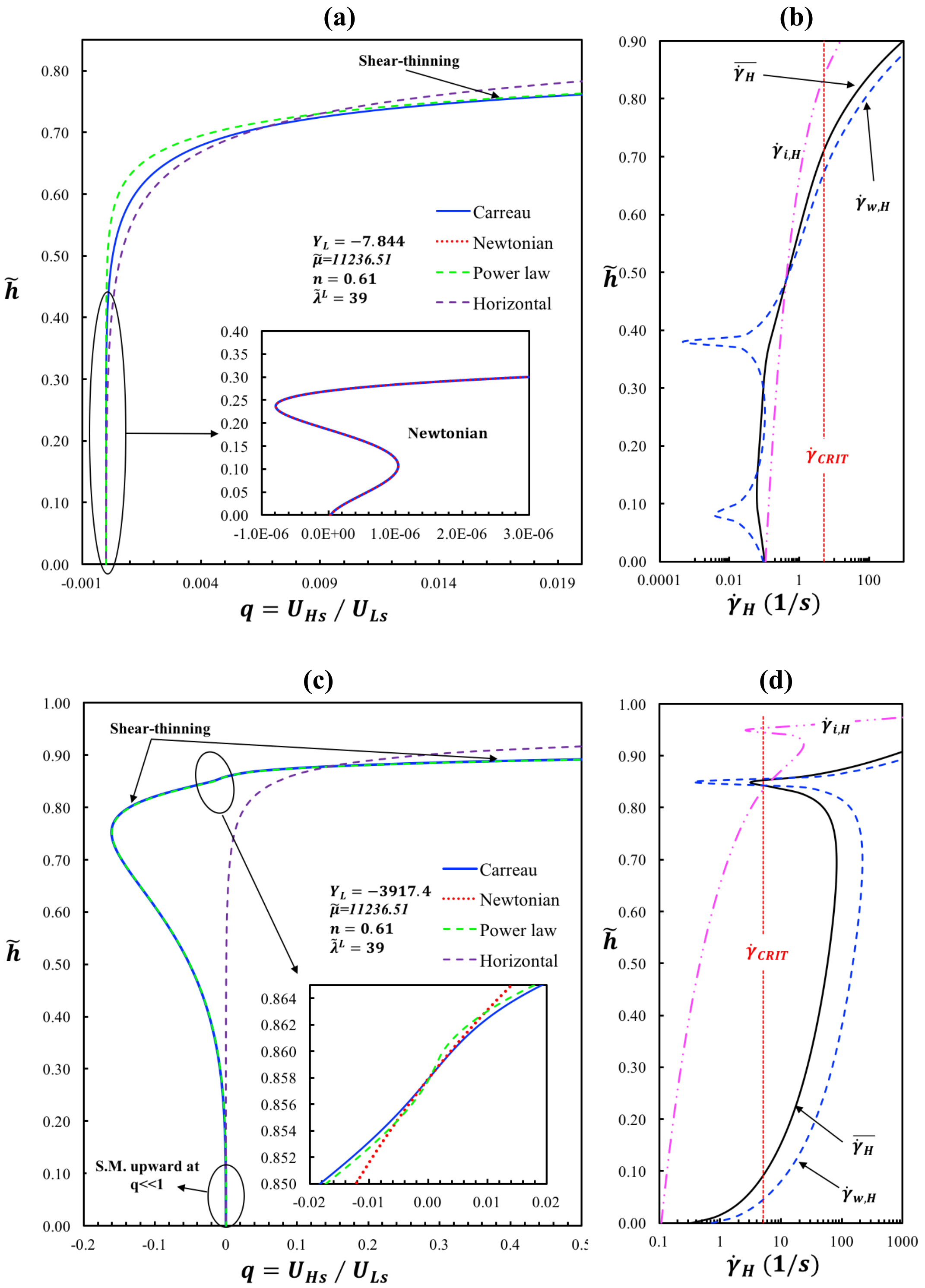

**Figure 15**. (a,c) Exact holdup curve for upward inclined gas/shear-thinning fluid stratified flow (Carreau fluid); the asymptotic hold-up curves (Newtonian zero-viscosity and power-law behaviour) and the holdup curve corresponding to horizontal flow are also plotted for two different channel inclinations. (b,d) Shear rate at the wall and at the interface, average shear rate of the shear-thinning fluid as a function of the holdup are plotted also. The combinations of nondimensional parameters corresponds to a 2 cm height channel, $U_{LS} = 0.01$ m/s, the inclination angle is $(a)\beta = 0.1°$ (upward), (b) $\beta = 5°$ (upward), and the shear-thinning fluid is CMC05.

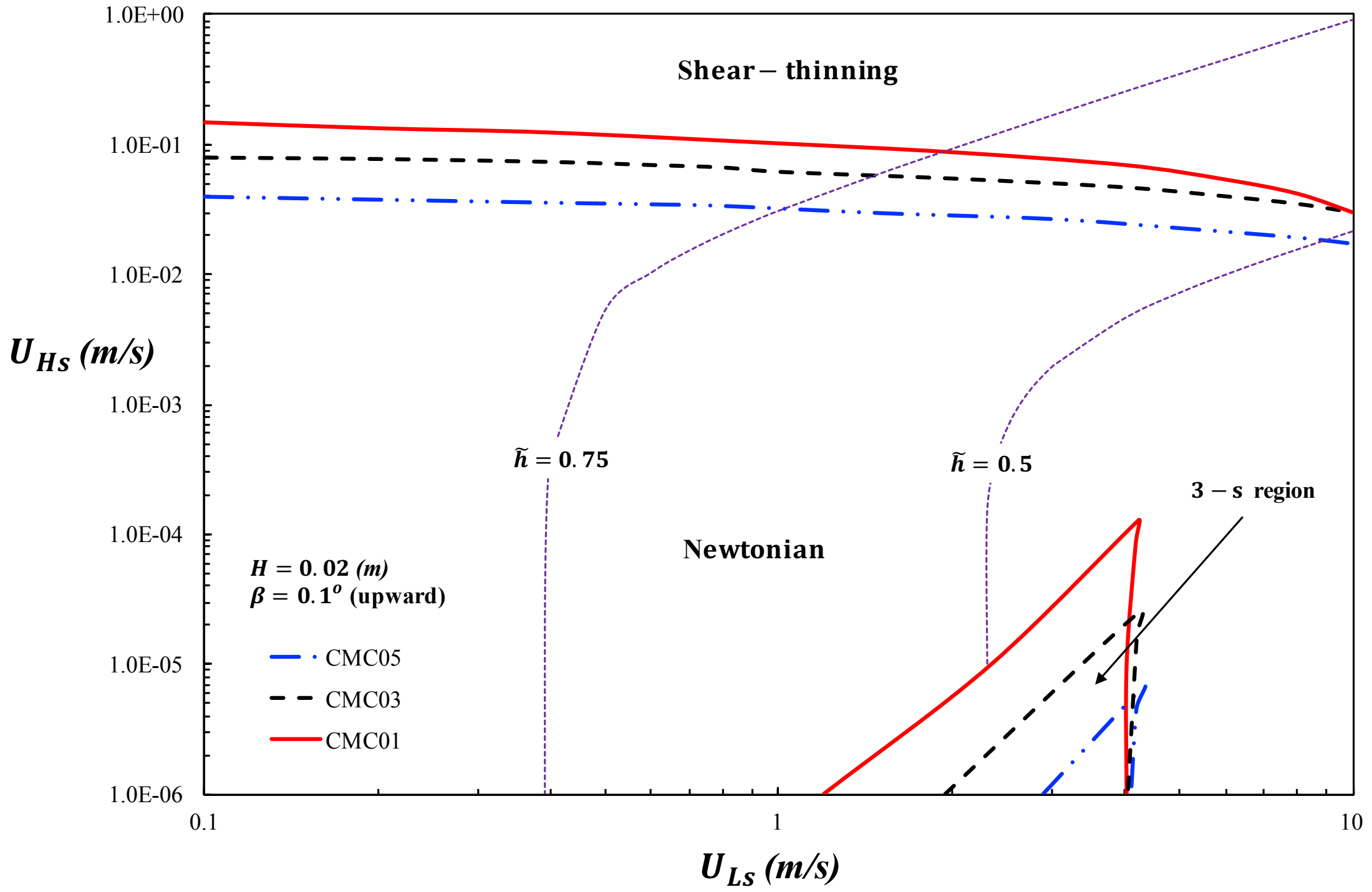

**Figure 16.** Rheological regions, asymptotic boundary, and triple solution region for upward inclined gas/shear-thinning fluid flows for different shear-thinning solutions. Iso-holdup (violet) for the CMC03 solution are also plotted.

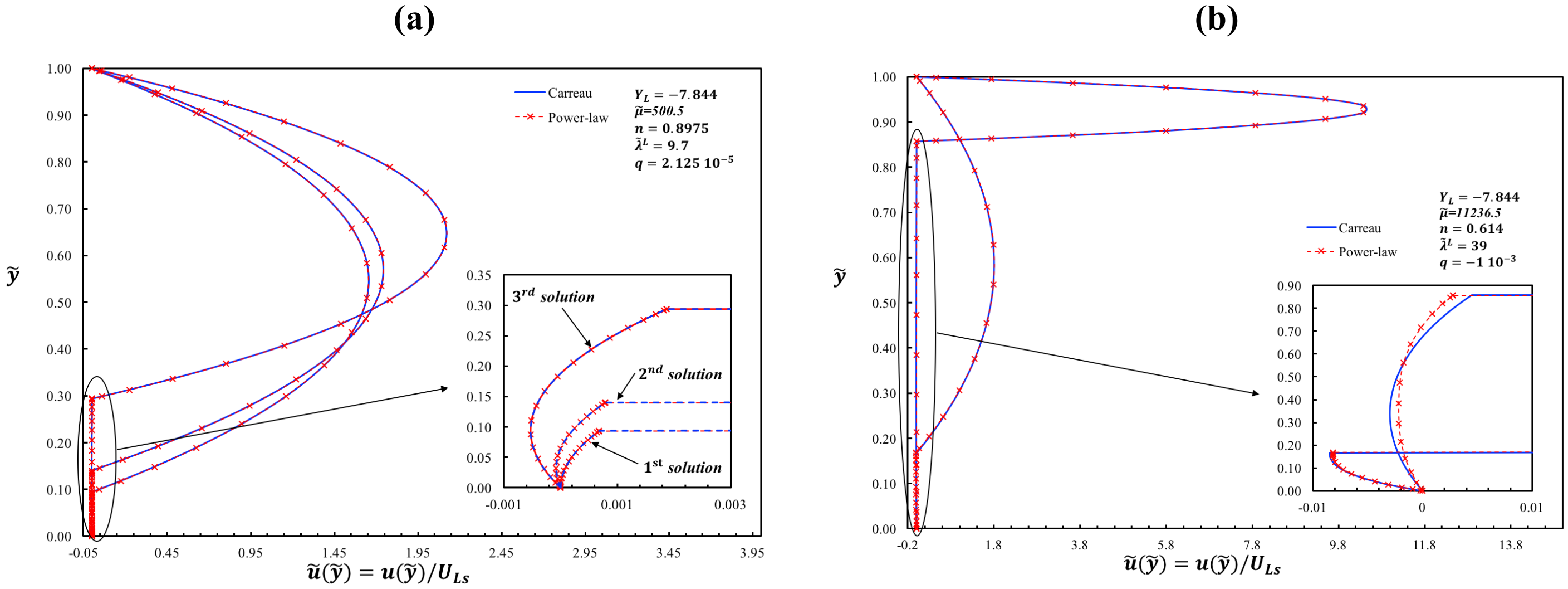

**Figure 17.** Velocity profile associated to triple solution in concurrent upward inclined, (a), and counter-current upward inclined, (b), for gas/shear-thinning fluid stratified flows. The exact (Careau fluid) solution, the power law and zero-viscosity Newtonian asymptotic solutions are plotted. The combinations of nondimensional parameters corresponds to a 2 cm height channel, $|U_{LS}| = 4$ m/s, the inclination angle is (a) $\beta = 0.1°$ (upward) and (b) $\beta = 5°$ (upward); the shear-thinning fluid is (a) CMC01 and (b) CMC05.

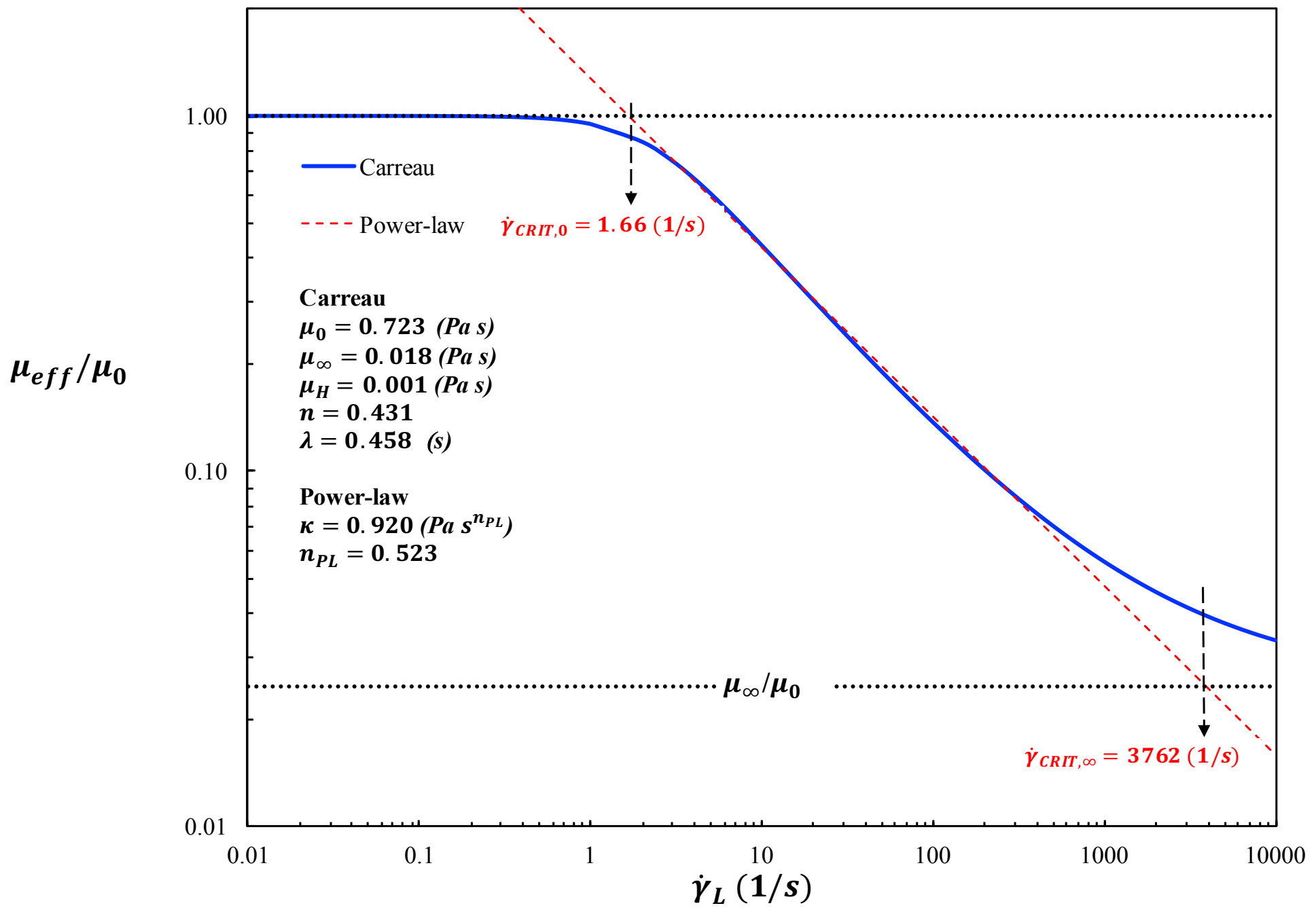

**Figure 18**. Viscosity curves for the shear-thinning emulsion by Pratal et al. (1997). $\dot{\gamma}_{CRIT,0}$ is the shear rate which corresponds to the intersection of the zero-shear-rate viscosity and the power-law model, while $\dot{\gamma}_{CRIT,\infty}$ indicates the transition to the infinity-shear-rate-viscosity behaviour.

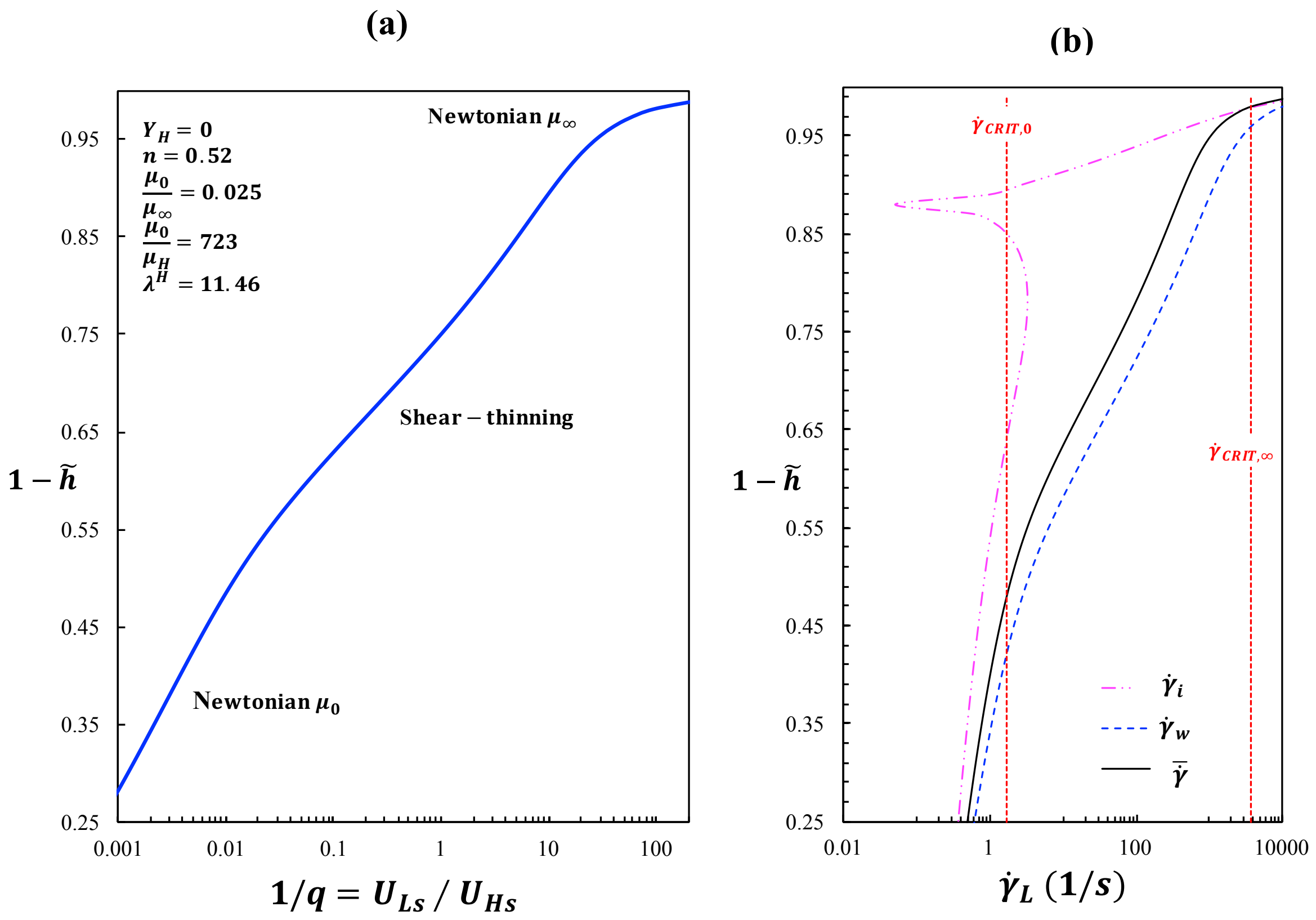

**Figure 19**. (a)Exact holdup curve for horizontal shear-thinning emulsion/water stratified flow (b) Shear rate at the wall and at the interface, average shear rate of the shear-thinning fluid as a function of the holdup.

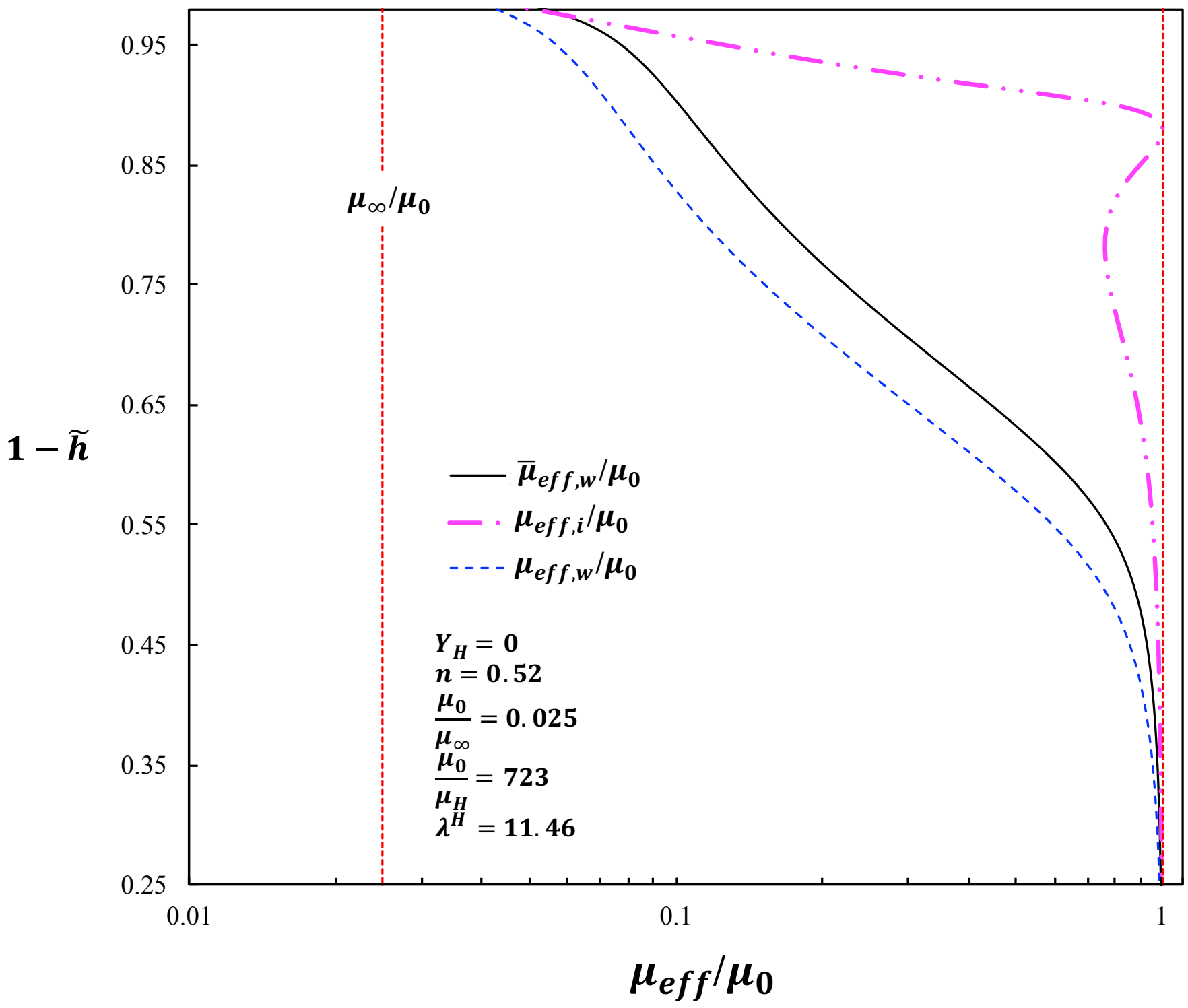

**Figure 20**. Trends of effective viscosity at the wall and at the interface, the average one for the shear-thinning fluid as a function of the holdup. The combinations of nondimensional parameters corresponds to a 2 cm height channel, $U_{HS} = 0.5$ m/s, horizontal flow and the shear-thinning fluid is the emulsion of Table 1.

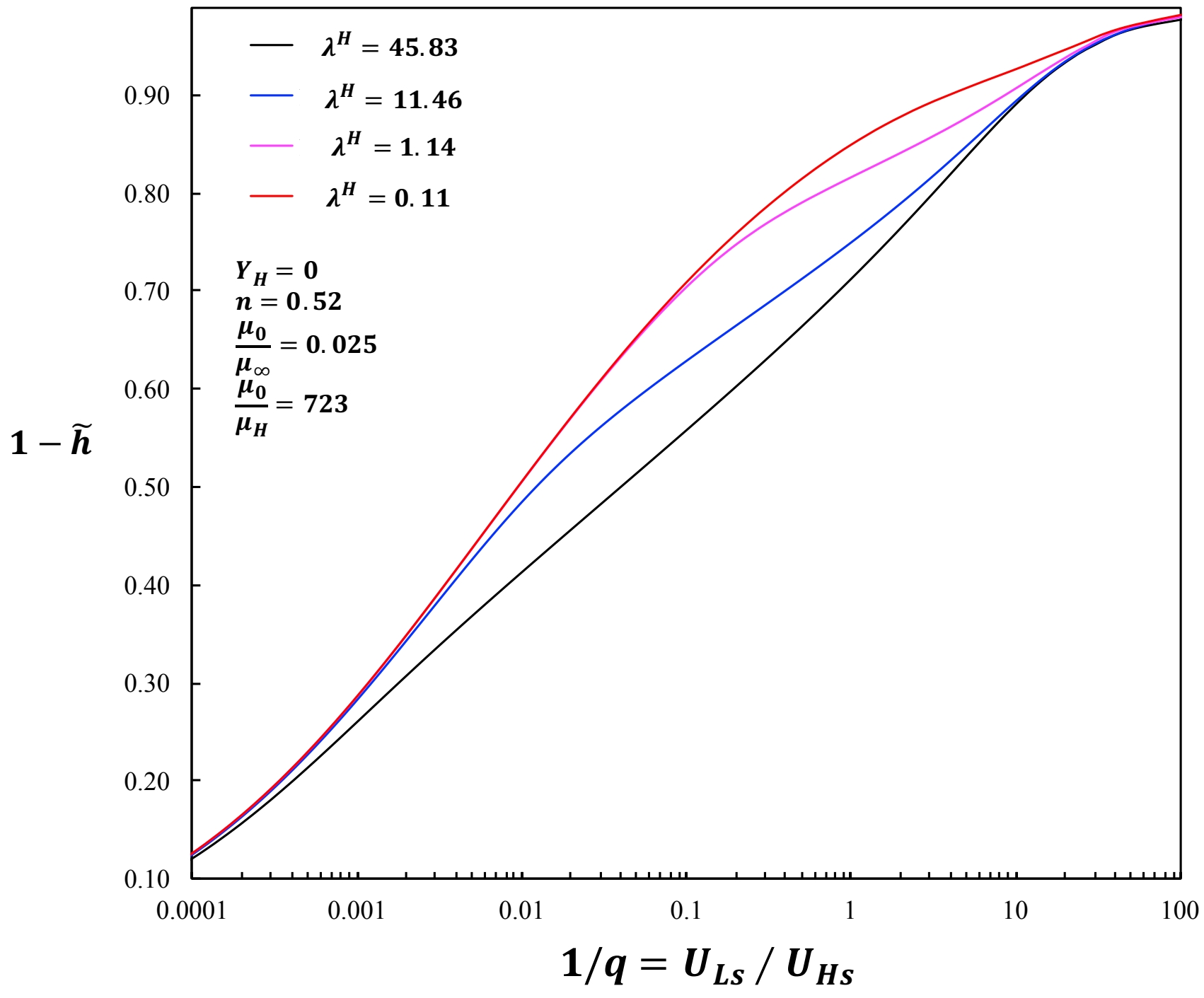

**Figure 21**. Trends of the holdup curve for different values of the dimensionless time constant for horizontal water/shear-thinning emulsion stratified flow.

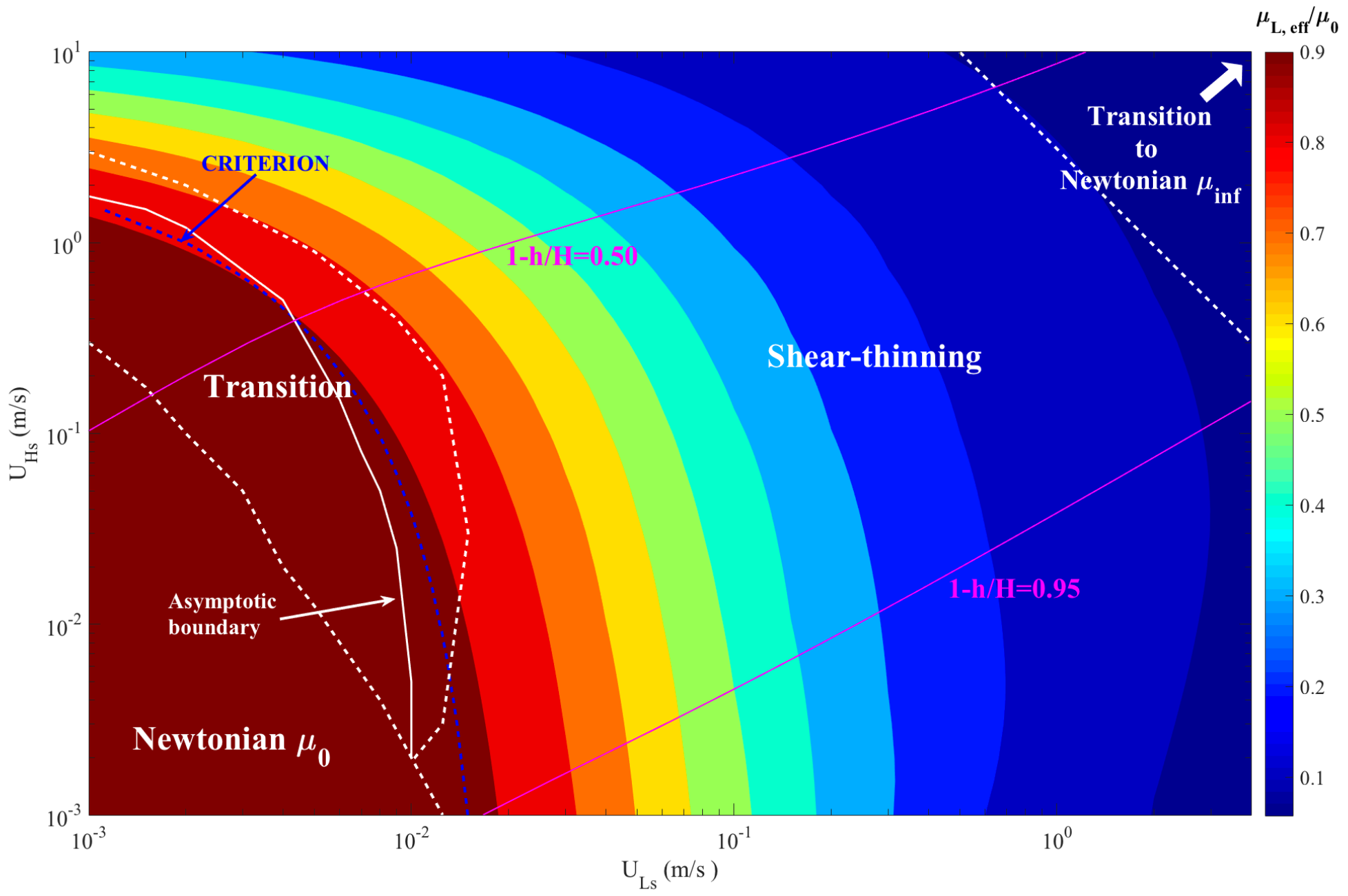

**Figure 22**. Flow maps for horizontal shear-thinning waxy oil/water stratified flow: water is the heavy phase and the emulsion is the light one. The contours of effective viscosity, the rheological regions (white), the iso-holdup curves (pink), and the asymptotic boundary (solid white line), which corresponds to the CRIT conditions of Fig. 17(b), are plotted. The CRITERION line (dashed blue) refers to the switch between the zero viscosity Newtonian and the power-law predicted by the two-fluid model, see Section 4.2.

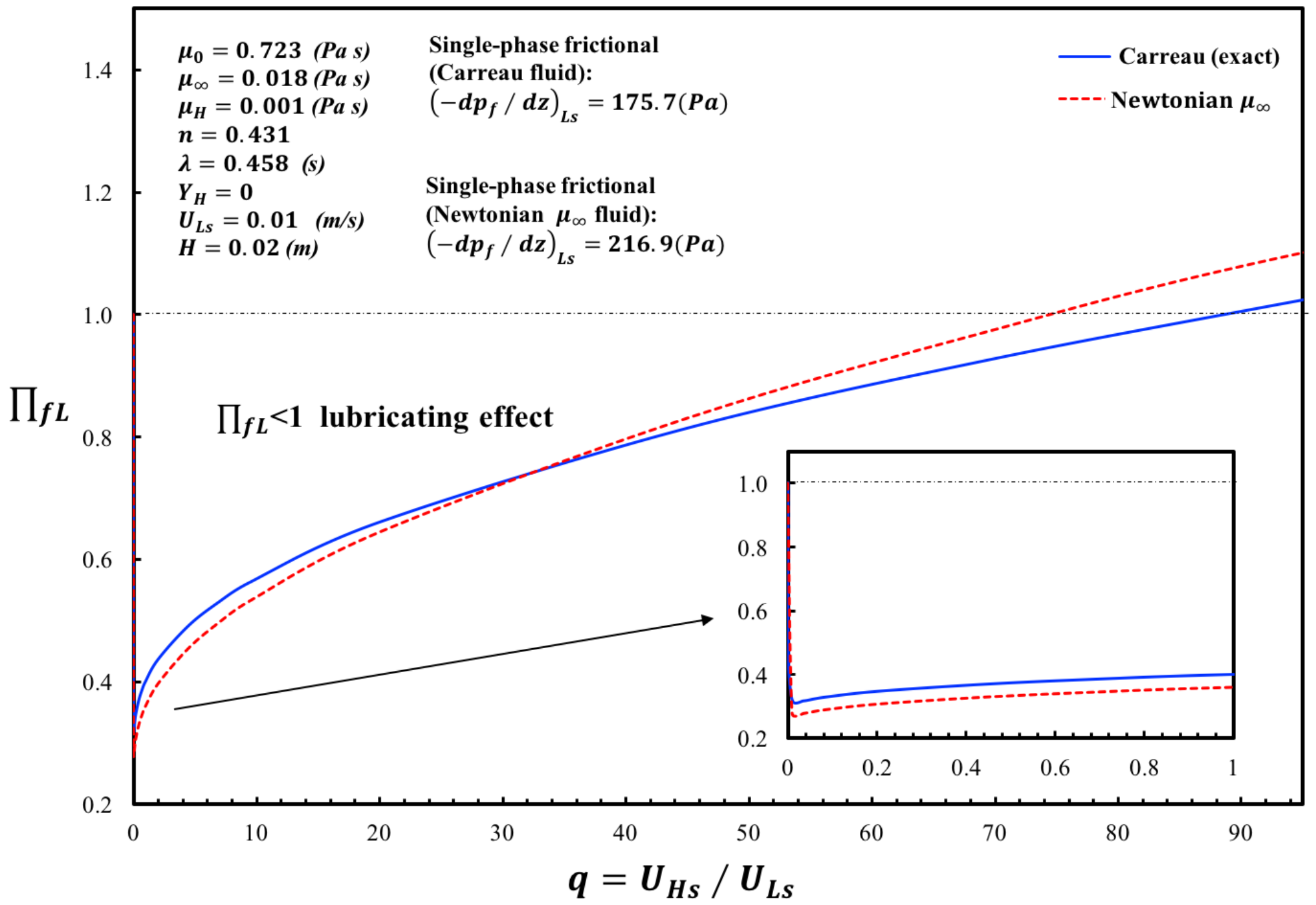

**Figure 23.** Frictional pressure gradient normalized by the single phase flow of the shear-thinning (Carreau) emulsion in order to show the lubrication effect by the addition of a heavy Newtonian phase (e.g. water). The case of a Newtonian fluid with zero-shear-rate viscosity is also shown to see the shear-thinning effect on the lubrication.

(a)

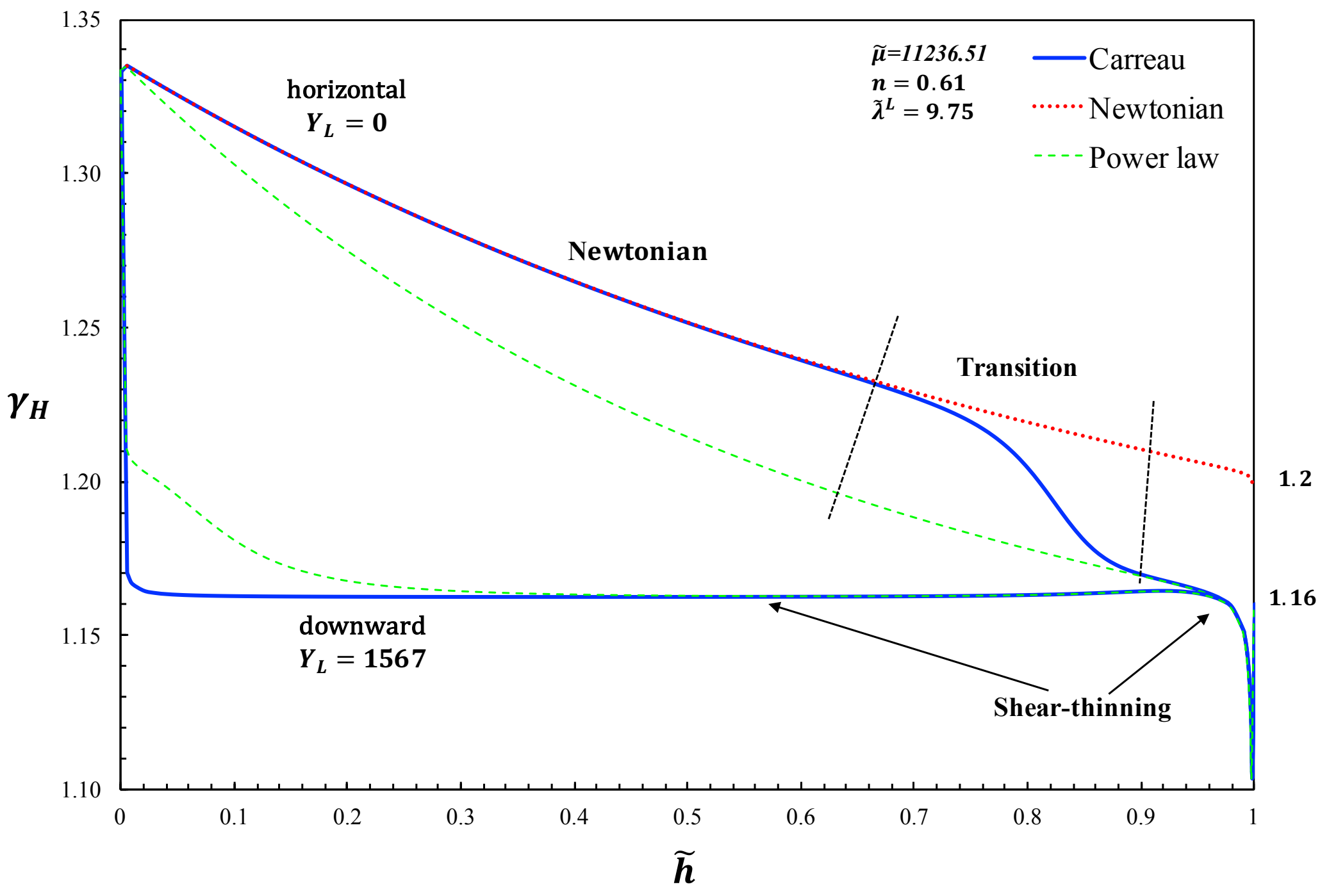

(b)

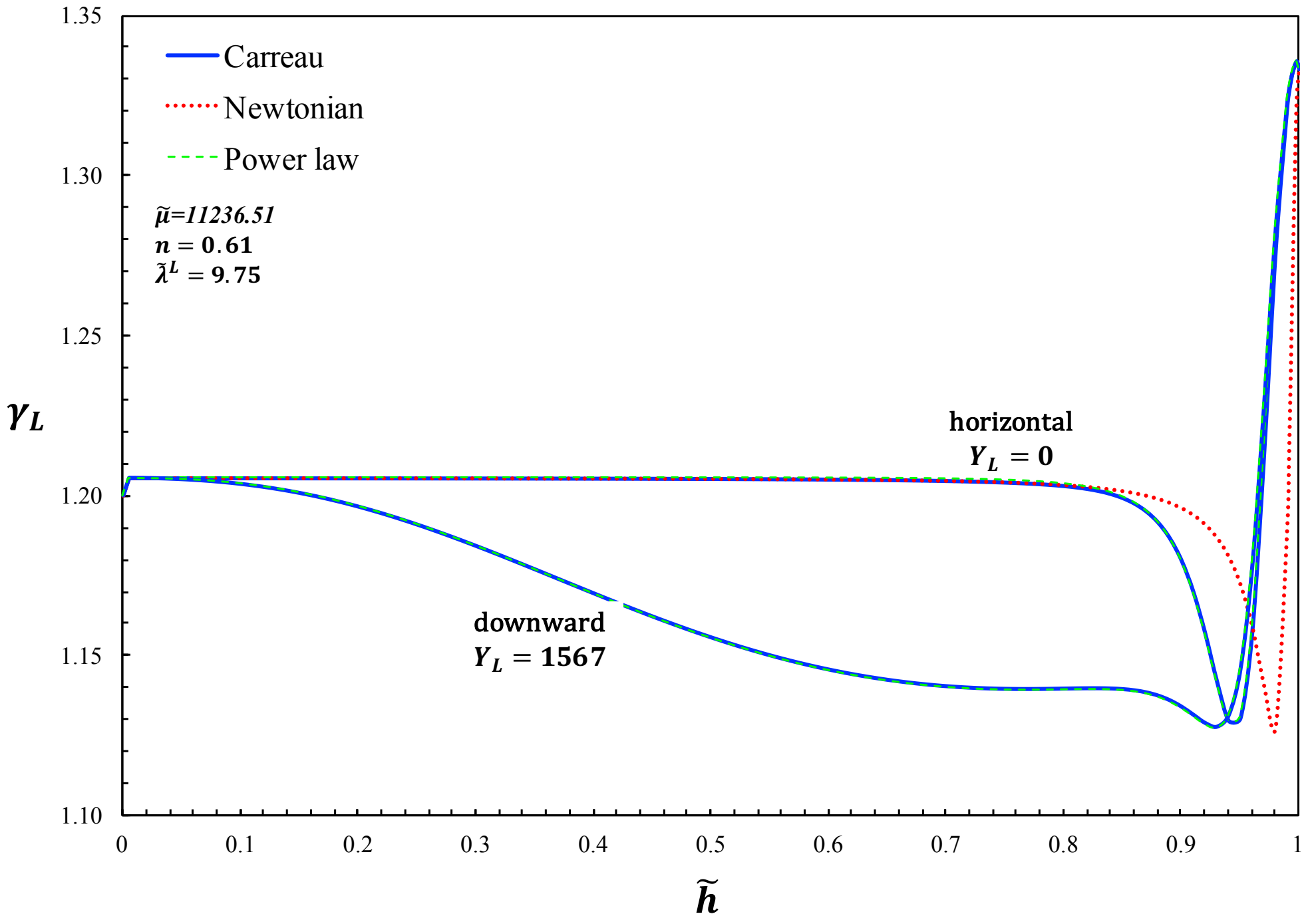

**Figure 24.** Shapes factors for horizontal and downward inclined gas/shear-thinning fluid stratified flows: the shapes factors trends corresponding to the asymptotic zero-viscosity Newtonian and power-law are also plotted. The combinations of nondimensional parameters corresponds to a 2 cm height channel, $U_{LS} = 1$ m/s, and the shear-thinning fluid is CMC05.

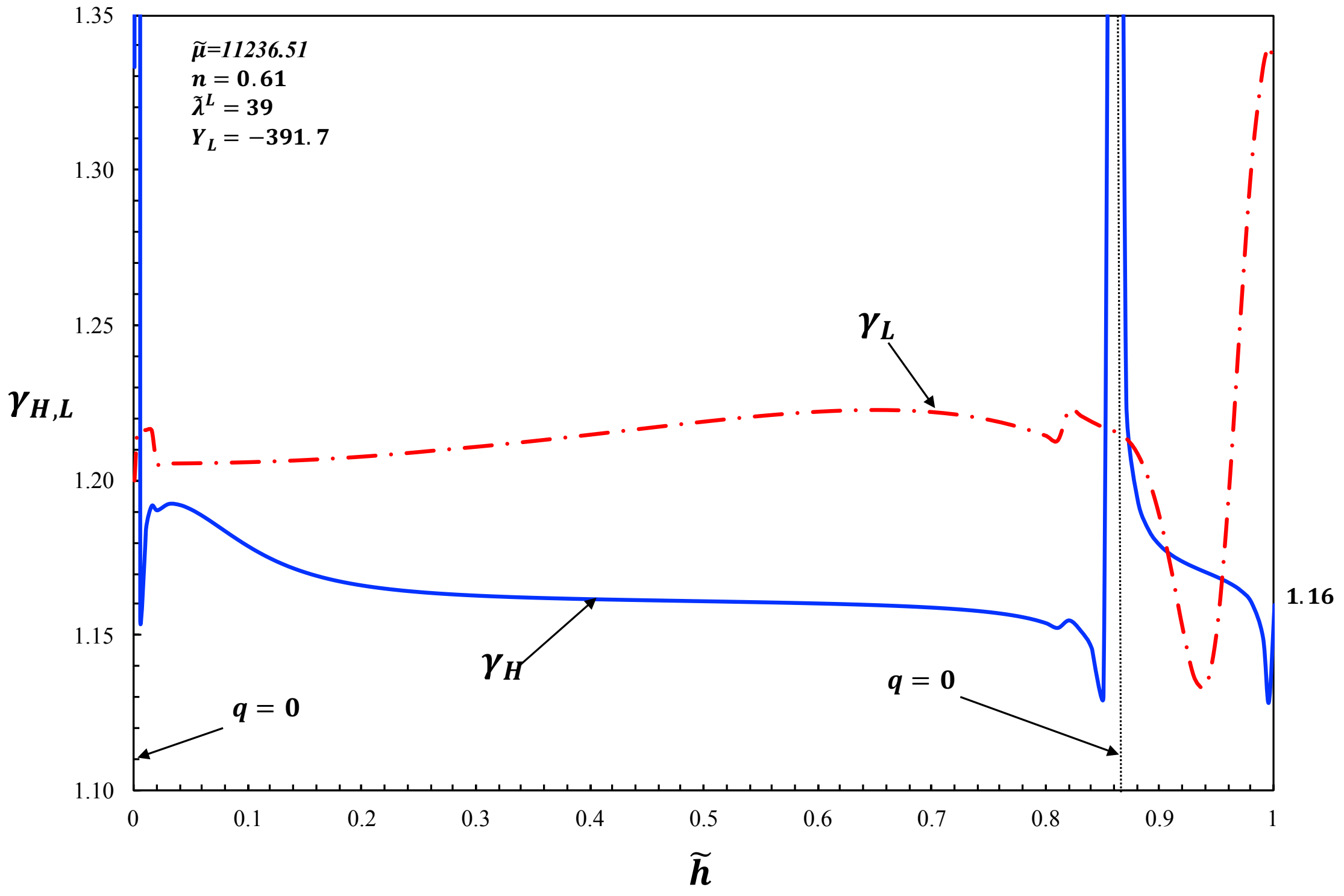

**Figure 25**. Shapes factors for upward inclined gas/shear-thinning fluid stratified flows. The conditions corresponding to the approaching of the recirculating points ($q = 0, \gamma_H \to \infty$) are presented. The combinations of nondimensional parameters corresponds to a 2 cm height channel, $|U_{LS|} = 4$ m/s, $\beta = 5°$ (upward), and the shear-thinning fluid is CMC05.

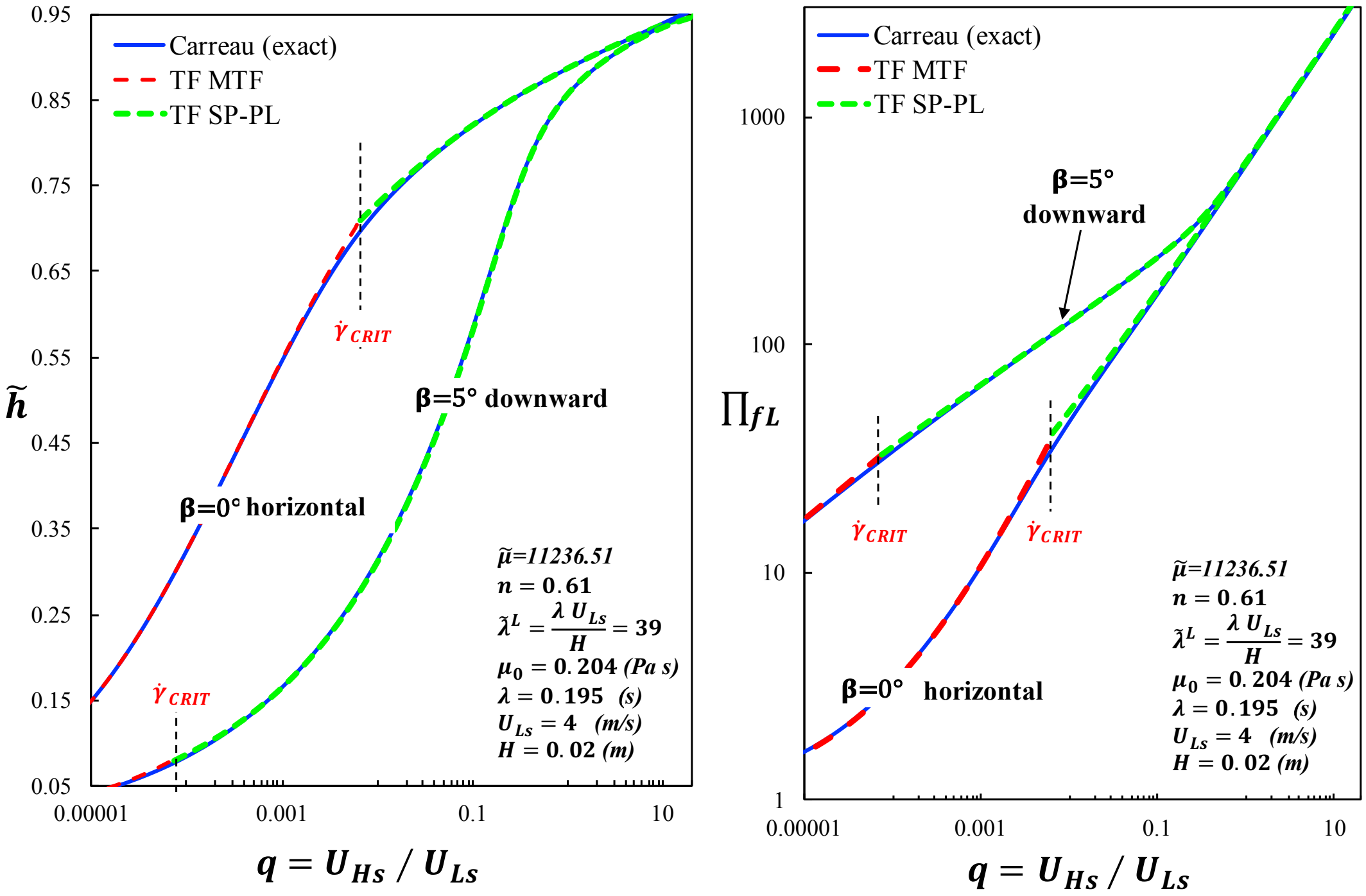

**Figure 26.** Comparison between the exact solution and the two-fluid model predictions using the proposed algorithm for horizontal and downward inclined gas/shear-thinning fluid flows: (a) Holdup (b) frictional pressure gradient.

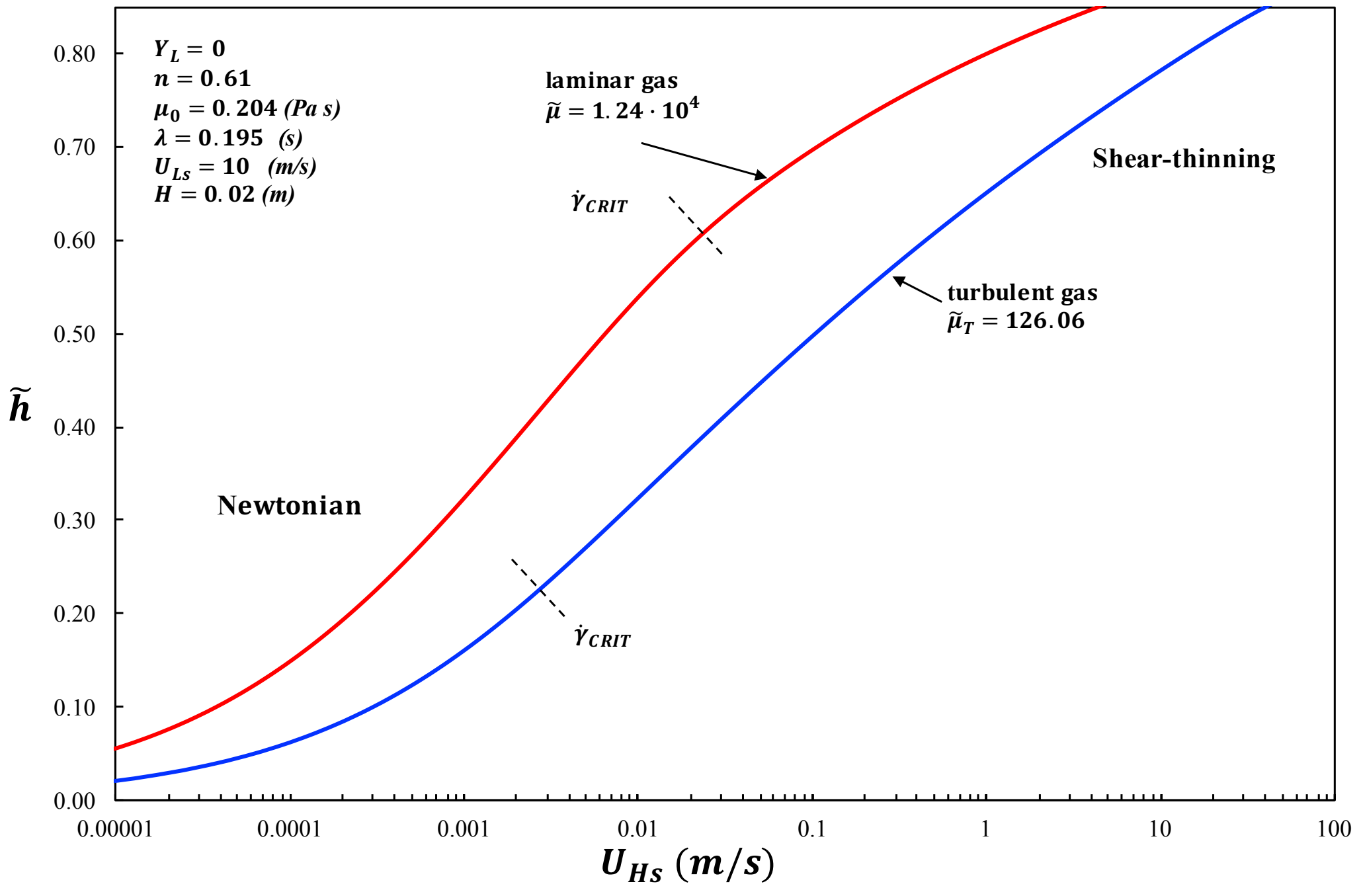

**Figure 27.** Holdup curves for horizontal laminar gas/laminar shear-thinning liquid and turbulent gas/laminar shear-thinning liquid stratified flow.